\documentclass{emulateapj}
\usepackage{url}
\usepackage{subfigure}
\usepackage{longtable}
\usepackage{amsmath}
\usepackage{graphicx}
\usepackage{subfigure}
\usepackage{longtable}
\usepackage{txfonts}
\usepackage{graphicx}
\usepackage{natbib}
\usepackage[dvipdfm]{hyperref}

\RequirePackage{color}

\begin{document}

\title{Exploring morphological correlations among H$_{2}$CO, $^{12}$CO, MSX and continuum
mappings}

\author{Chuan Peng Zhang\altaffilmark{1,2}, Jarken
Esimbek\altaffilmark{1,3}, Jian Jun Zhou\altaffilmark{1,3}, Gang
Wu\altaffilmark{1,3} \& Zhi Mao Du\altaffilmark{1,2}}

\altaffiltext{1}{Xinjiang Astronomical Observatory, Chinese Academy
of Sciences, Urumqi, 830011, China; zcp0507@gmail.com.}
\altaffiltext{2}{Graduate University of the Chinese Academy of
Sciences, Beijing, 100080, China.} \altaffiltext{3}{Key Laboratory
of Radio Astronomy, Chinese Academy of Sciences, Urumqi, 830011,
China.}

\begin{abstract}
There are relatively few H$_{2}$CO mappings of large-area giant
molecular cloud (GMCs). H$_{2}$CO absorption lines are good tracers
for low-temperature molecular clouds towards star formation regions.
Thus, the aim of the study was to identify H$_{2}$CO distributions
in ambient molecular clouds. We investigated morphologic relations
among 6-cm continuum brightness temperature (CBT) data and H$_{2}$CO
($1_{11}-1_{10}$; Nanshan 25-m radio telescope), $^{12}$CO (1--0;
1.2-m CfA telescope) and midcourse space experiment (MSX) data, and
considered the impact of background components on foreground clouds.
We report simultaneous 6-cm H$_{2}$CO absorption lines and
H110$\alpha$ radio recombination line observations and give several
large-area mappings at 4.8 GHz toward W49 ($50'\times50'$), W3
($70'\times90'$), DR21/W75 ($60'\times90'$) and NGC2024/NGC2023
($50'\times100'$) GMCs. By superimposing H$_{2}$CO and $^{12}$CO
contours onto the MSX color map, we can compare correlations. The
resolution for H$_{2}$CO, $^{12}$CO and MSX data was
$\thicksim$10$'$, $\thicksim$8$'$ and $\thicksim$18.3$''$,
respectively. Comparison of H$_{2}$CO and $^{12}$CO contours,
8.28-$\mu$ m MSX colorscale and CBT data revealed great
morphological correlation in the large area, although there are some
discrepancies between $^{12}$CO and H$_{2}$CO peaks in small areas.
The NGC2024/NGC2023 GMC is a large area of HII regions with a high
CBT, but a H$_{2}$CO cloud to the north is possible against the
cosmic microwave background. A statistical diagram shows that
85.21\% of H$_{2}$CO absorption lines are distributed in the
intensity range from $-1.0$ to 0 Jy and the $\Delta V$ range from
1.206 to 5 km s$^{-1}$.
\end{abstract}

\keywords{formation; massive; clouds; HII regions; imaging;
individual (W49, W3, DR21/W75 \& NGC2024/NGC2023)}


\section{Introduction} \label{introduction}

Absorption lines for formaldehyde (H$_{2}$CO; $J_{KaKc}$ =
1$_{11}-1_{10}$; $\nu_{o}$ = 4829.659 4MHz), discovered in the
interstellar medium by \citet{sny69}, are commonly detected toward
star formation regions. H$_{2}$CO is a slightly asymmetric rotor
molecule and is inherently sensitive to kinetic temperature.
H$_{2}$CO is an accurate probe of physical conditions in dense
molecular clouds \citep{man93}. Anomalous absorption lines can be
seen against the 2.7 K cosmic microwave background
\citep[CMB;][]{pal69} and detected in dark clouds. Absorption is
strongest at high density and temperature owing to the collisional
pumping mechanism. A survey by \citet{dow80} between $l = 0^\circ$
and 60$^\circ$, $b = \pm1^\circ$ suggested that HII regions are
associated with H$_{2}$CO. 12 bright galactic HII regions and two
dark clouds (W3, W3(OH), NGC2024, W31, W33, M17, W43, W49A, W51A,
W51B, K3-50 and DR21/W75; NGC2264 and Heiles cloud 2) have been
mapped at an angular resolution of 2.6$'$ using a 100-m telescope
\citep{bie82}, but their mapping areas are smaller than ours. Toward
L1551, observations made using the Arecibo Telescope showed that
H$_{2}$CO lines not only trace the outer envelope or quiescent state
of molecular clouds, but also provide kinematic information on star
formation \citep{ara06}. \citet{pie88} presented a first extensive
survey of the Cyg X-region using H$_{2}$CO and H110$\alpha$ lines as
measured with the 100 m-RT Effelsberg telescope. Heiles clouds 1 and
2 and Lynds cloud L134 were mapped using the Onsala 25-m telescope
and the results show that H$_{2}$CO contours generally coincide with
areas of optical obscuration \citep{sum75}. \citet{rod06} carried
out a blind search of anomalous H$_{2}$CO absorption lines in the
general direction of the galactic anticenter radiation using the
Onsala 25-m radio telescope. This large-area mapping shows that
H$_{2}$CO absorption and $^{12}$CO emission lines spatially coexist
and all H$_{2}$CO absorption features observed are associated with
known $^{12}$CO emission. In addition, \citet{rod07} mapped a large
region around L1204/S140 using H$_{2}$CO absorption line as measured
with the Onsala 25-m telescope and concluded that $^{12}$CO emission
and H$_{2}$CO absorption lines correlate fairly well, both
qualitatively and quantitatively. Nevertheless, large-scale mappings
of star formation regions using H$_{2}$CO absorption lines,
especially for large-area GMCs, are relatively few.

\citet {mck07} offered a basic description of turbulence, magnetic
fields and self-gravity for the key dynamic processes involved in
star formation. These three mechanisms influence the systematic
velocity and broaden the line width. In general, most massive star
formation regions deposit ultra-compact HII (UCHII) regions, which
are dense and compact bubbles of photoionized gas of less than 0.1
pc in diameter and are surrounded by massive young stellar objects
\citep[YSOs;][]{tho06}. Measurement of H110$\alpha$ radio
recombination lines (RRLs; $\nu_{o}$ = 4874.157 MHz) and H$_{2}$CO
absorption can be used to determine the kinematic distance of UCHII
regions \citep[e.g.][]{ara02,wat03}. To discern the overall
structure of GMCs, correlations among various molecular clouds
should be considered and large-scale surveys are necessary.

In this paper, we report large-scale mappings of W49, W3, DR21/W75
and NGC2024/NGC2023 GMCs, almost all of which have a thermal
background. We compare the relations among H$_{2}$CO, $^{12}$CO and
MSX data and the continuum brightness temperature (CBT). The
remainder of the paper is organized as follows. Section
\ref{ob-and-data} describes the observations, data production and
relevant parameters. In Section \ref{data-dscption-analy}, we
presents mappings for H$_{2}$CO, $^{12}$CO, MSX and CBT data. In
addition, we report some associations among $^{12}$CO, H$_{2}$CO and
continuum spectra, and define ranges for relations between line
width and intensity. Finally, conclusions are presented in Section
\ref{summa}.

\section{Observations and Data} \label{ob-and-data}

\subsection{Observations} \label{obser}

Observations were made discontinuously from October 2009 to January
2011 using the Urumqi Nanshan 25-m telescope, which affords a
half-power beam size (HPBW) of approximately 10$'$ at 4.8 GHz.
H$_{2}$CO absorption and H110$\alpha$ RRL spectra were
simultaneously obtained using a 4096-channel digital autocorrelation
spectrometer at a bandwidth of 80 MHz using position switching. The
central velocity was $V_{LSR}$ = 0.0 km s$^{-1}$, the frequency was
$F_{center}$ = 4851.9102 MHz, and the velocity channel width was
1.206 km s$^{-1}$. To achieve a better signal-to-noise ratio (S/N),
the ON-source integration time was tens of minutes to up to several
hours. CBT data at 4.8 GHz were processed with a bandwidth of about
400 MHz. The CBT error was approximately 1\% and sensitivity was
approximately 75 mJy. The system temperature during observations was
23 K. A diode noise source was used to calibrate the spectra and the
flux error was 15\%. The pointing accuracy was better than 15$''$
for all observations and the antenna efficiency was 62\%. The DPFU
(degrees per flux unit) value was 0.116 K Jy$^{-1}$.

\subsection{Data Reduction} \label{data-re}

Data reduction for H$_{2}$CO absorption lines and H110$\alpha$ RRLs
was performed using CLASS and GREG, which are part of the GILDAS
software developed by IRAM. Furthermore, 8.28-$\mu$ m mid-infrared
MSX archive data were handled using SAOImage DS9, Adobe Photoshop
7.0.1, Origin 8 and Adobe Illustrator CS software. Starlink software
was used to process $^{12}$CO cube data \citep{dam01} to obtain
parameters (integration intensity, velocity, intensity and line
width) for comparison with H$_{2}$CO parameters.

\subsection{Data Exhibition} \label{data-exhi}

MSX data traces star formation regions, and mainly traces the dust
continuum emission from core, so that we could compare the
distributional relation between H2CO and MSX data, and shed light on
the background environment for star formation regions, CBT data at
4.8 GHz and MSX data for the 8.28-$\mu$ m band were used as the
background for H$_{2}$CO contours. The correlation between MSX
colorscale images and H$_{2}$CO absorption contours is described in
Section \ref{data-dscption-analy}. The 2.6-mm $^{12}$CO data from
the 1.2-m CfA telescope \citep{dam01} were used as a background to
determine their relationship. The CO beam size was $\thicksim$8$'$,
which is comparable to that for H$_{2}$CO ($\thicksim$10$'$).

The optical depth, column density and mass for measured H$_{2}$CO
absorption lines were calculated in the usual way. According to
\citet{bie82}, the apparent optical depth is
\begin{eqnarray}
    \label{eq:tau}
    \tau_{app} = -ln[1+\frac{T_{L}}{T_{c}+T_{bgd}-T_{ex}}],
\end{eqnarray}
where $T_{L}$ is the line intensity measured, $T_{c}$ is the CBT and
$T_{bgd}$ is the 2.7 K cosmic background temperature. $T_{ex}$ is
the 1.7 K excitation temperature of the H$_{2}$CO $1_{10}-1_{11}$
transition. The H$_{2}$CO column density was calculated at the
$1_{11}$ level as:
\begin{eqnarray}
    \label{eq:nh2co}
    N(H_{2}CO) = 9.4 \times 10^{13} \tau_{app} \cdot \Delta V,
\end{eqnarray}
where $\Delta V$ is the H$_{2}$CO FWHM in km s$^{-1}$. Thus, the
column density is \citep{sco73}:
\begin{eqnarray}
    \label{eq:nh2}
    N(H_{2}) = 0.8 \times 10^{9} N(H_{2}CO).
\end{eqnarray}
Finally, we used the following equation \citep{poe83}
\begin{eqnarray}
    \label{eq:mh21}
    M(H_{2}) = S \cdot \overline{N(H_{2})} \cdot m_{H_{2}} \cdot r^{2}
\end{eqnarray}
to indirectly derive the clump's $H_{2}$-mass:
\begin{eqnarray}
    \label{eq:mh22}
    M(H_{2}) = 1.36 \times 10^{-21} \cdot \frac{S}{(arcmin)^{2}} \cdot
\overline{N(H_{2})} \cdot M_{\odot}  \cdot r^{2},
\end{eqnarray}
where $S$ is the observational area, $\overline{N(H_{2})}$ is the
average column density, $m_{H_{2}}$ is the mass of the
$H_{2}$-molecular and $r$ (kpc) is their distances.

\begin{figure}
\centering \subfigure[W49 GMC]{ \label{w49a}
\includegraphics[width=0.9\textwidth]{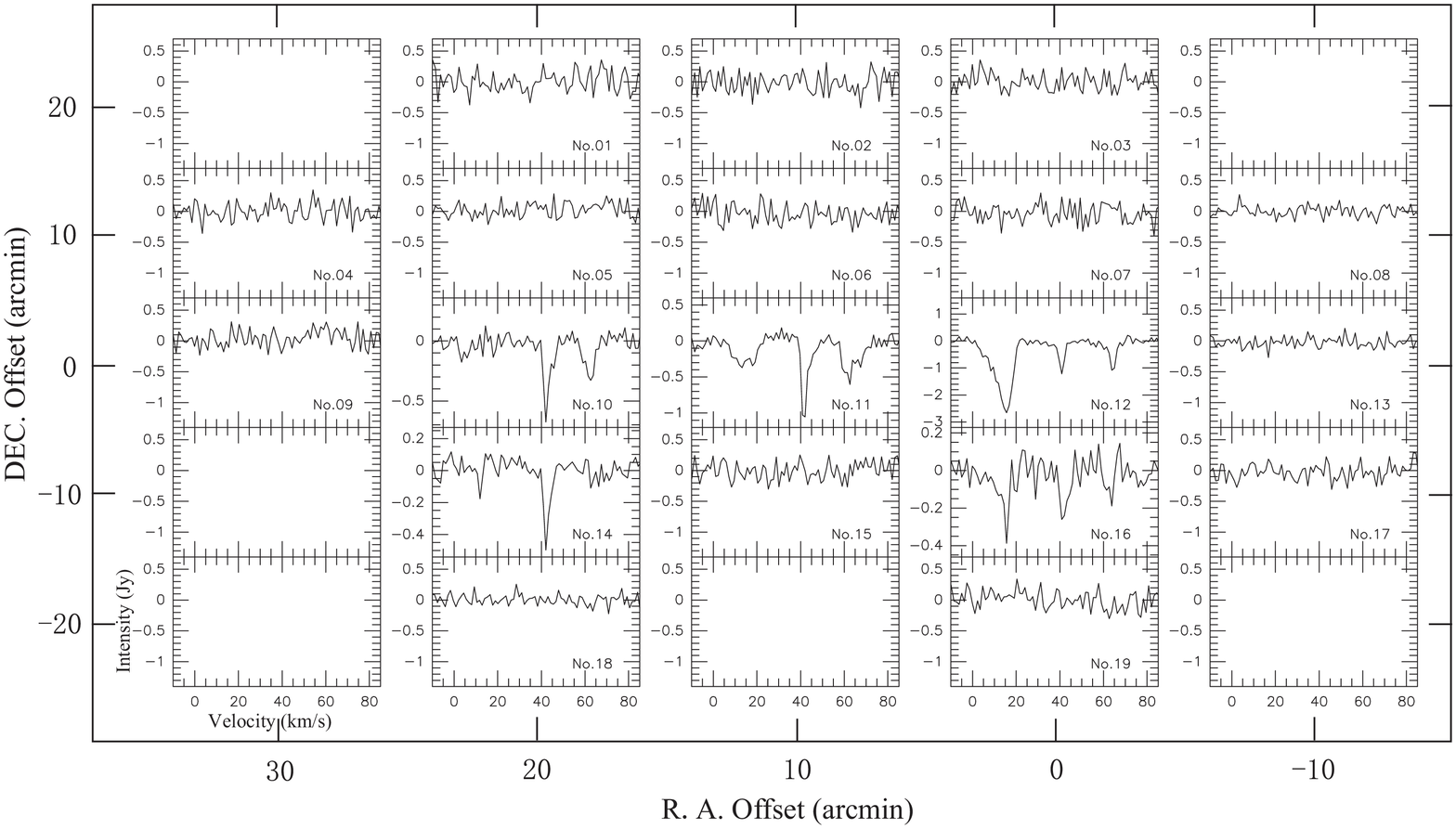}}
\hspace{1in} \subfigure[W49 GMC]{ \label{w49b}
\includegraphics[width=0.9\textwidth]{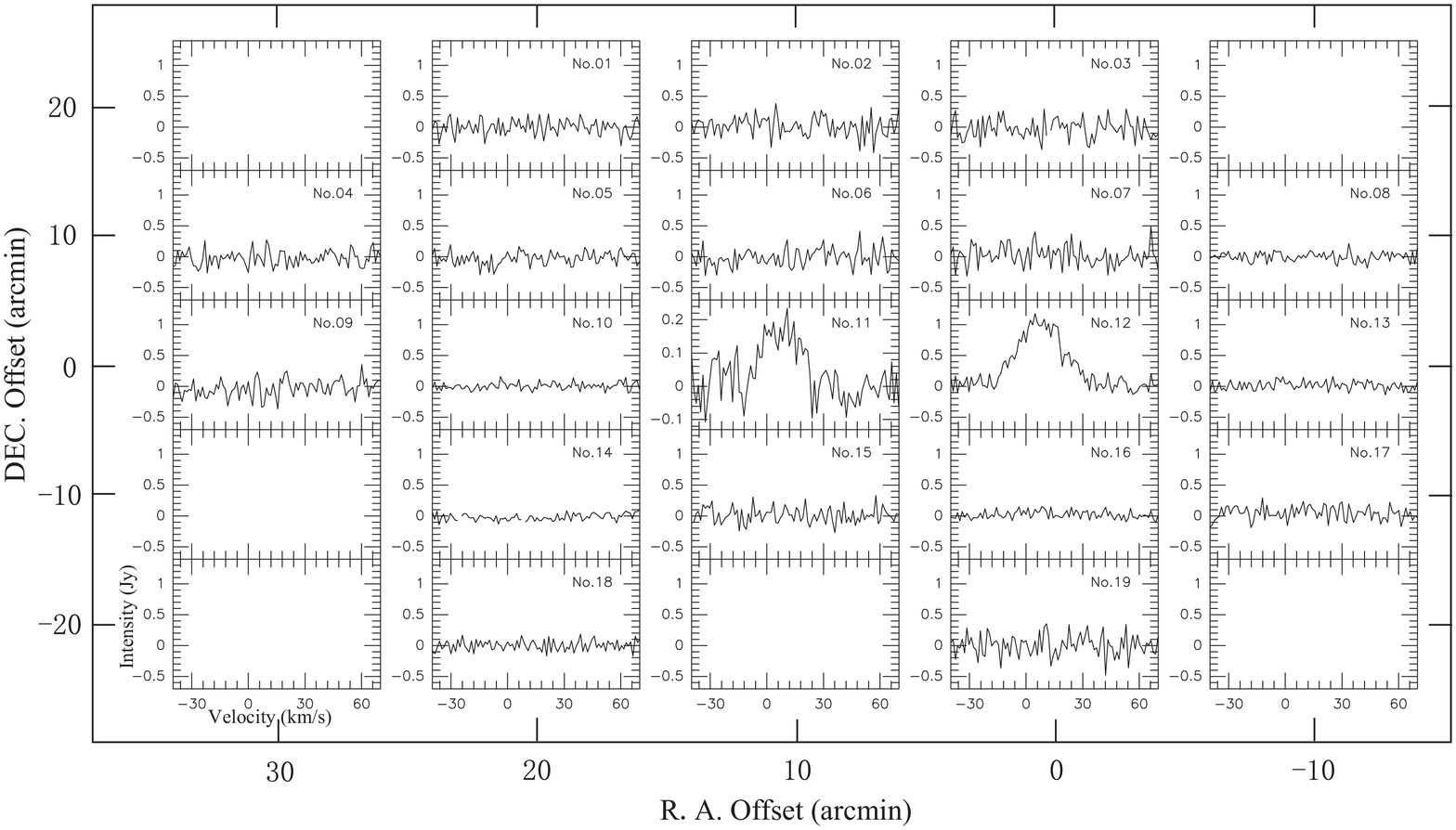}}
\caption{The spectral mosaic of (a) H$_{2}$CO absorption line and
(b) H110$\alpha$ RRL toward W49 GMC.}\label{w49ab}
\end{figure}

\begin{figure}
\centering \subfigure[W3 GMC]{\label{w3a}
\includegraphics[width=0.8\textwidth]{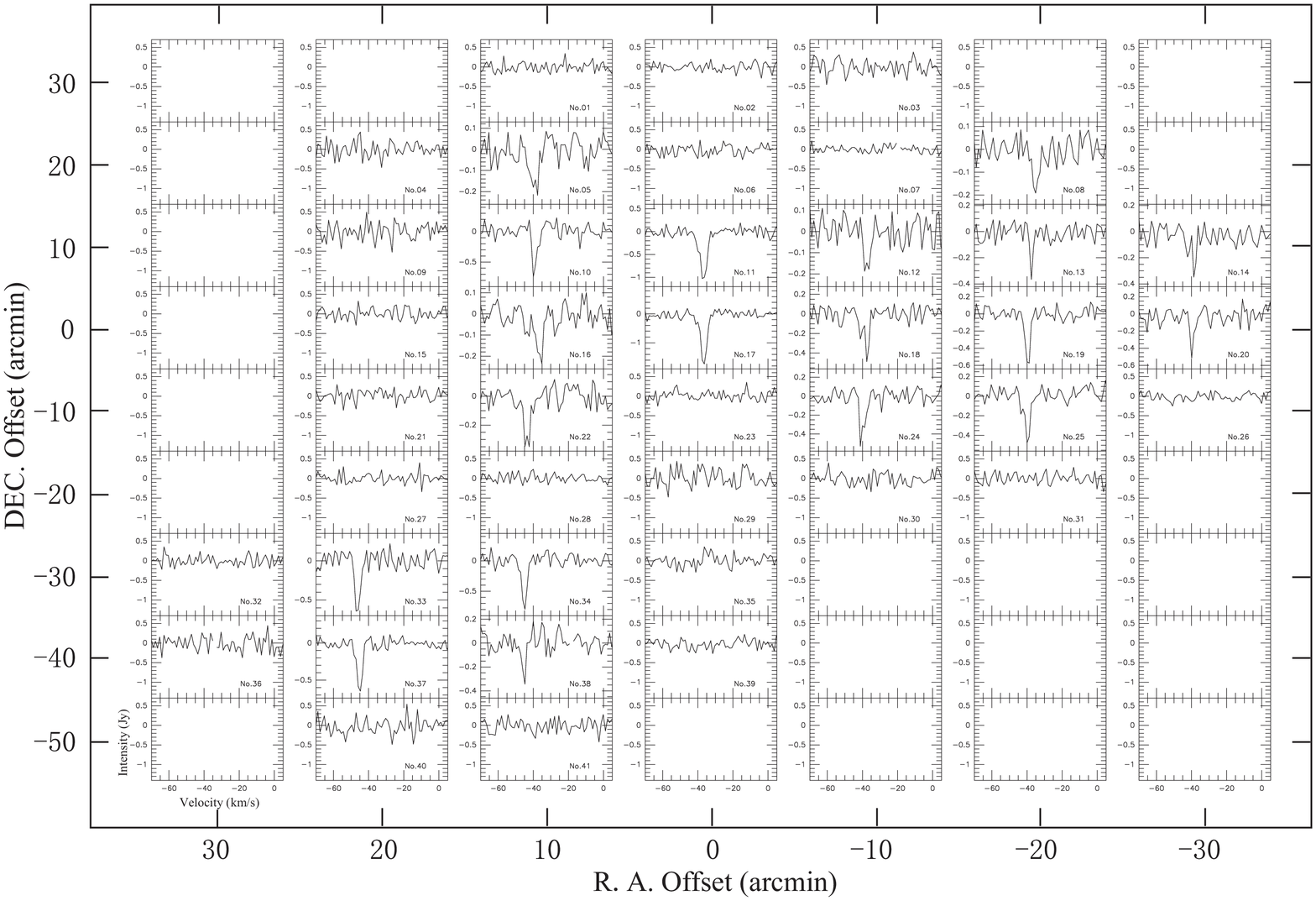}}
\hspace{1in} \subfigure[W3 GMC]{\label{w3b}
\includegraphics[width=0.8\textwidth]{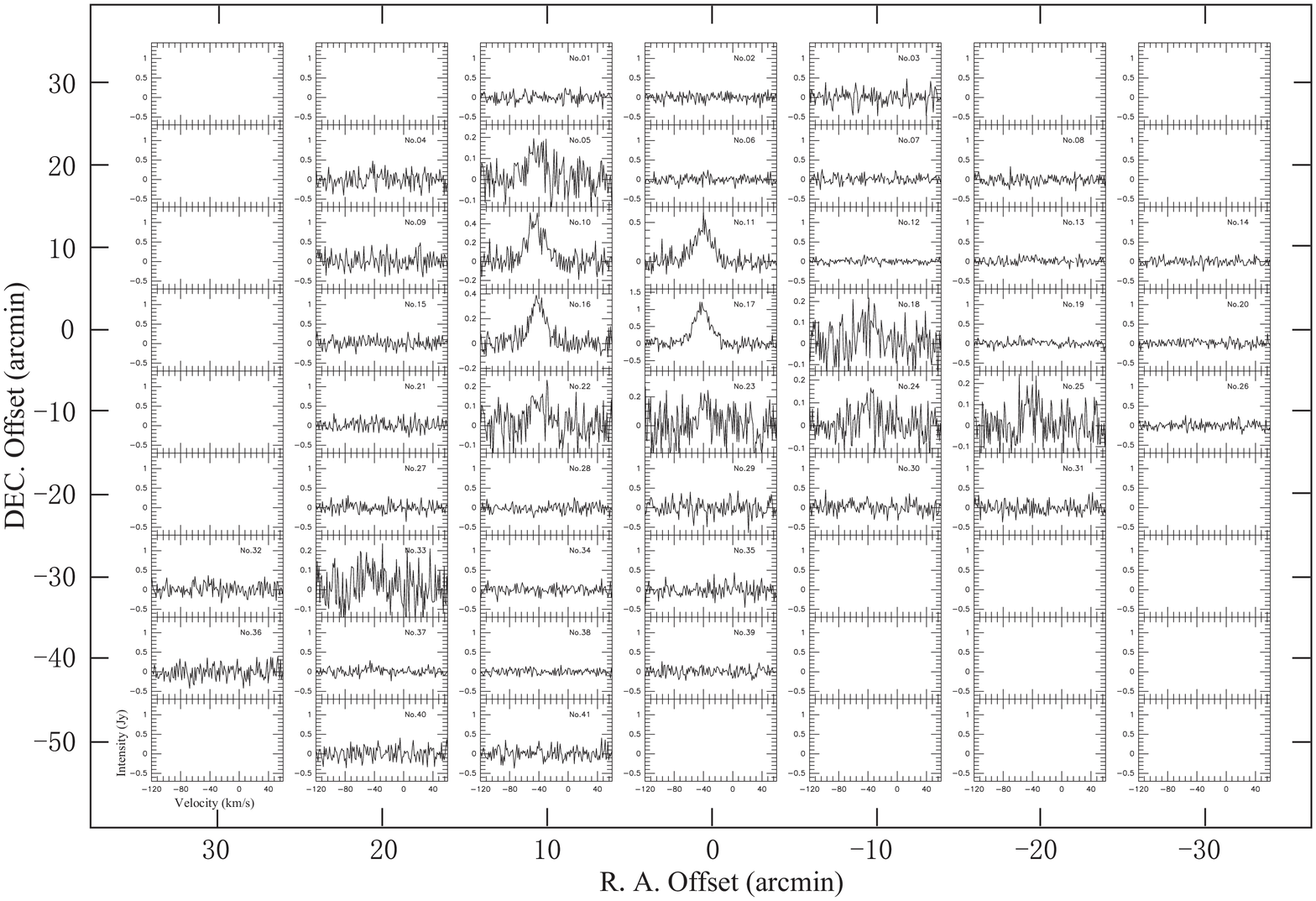}}
\caption{The spectral mosaic of (a) H$_{2}$CO absorption line and
(b) H110$\alpha$ RRL toward W3 GMC.}\label{w3ab}
\end{figure}

\begin{figure}
\centering \subfigure[DR21/W75 GMC]{\label{dr21a}
\includegraphics[width=0.8\textwidth]{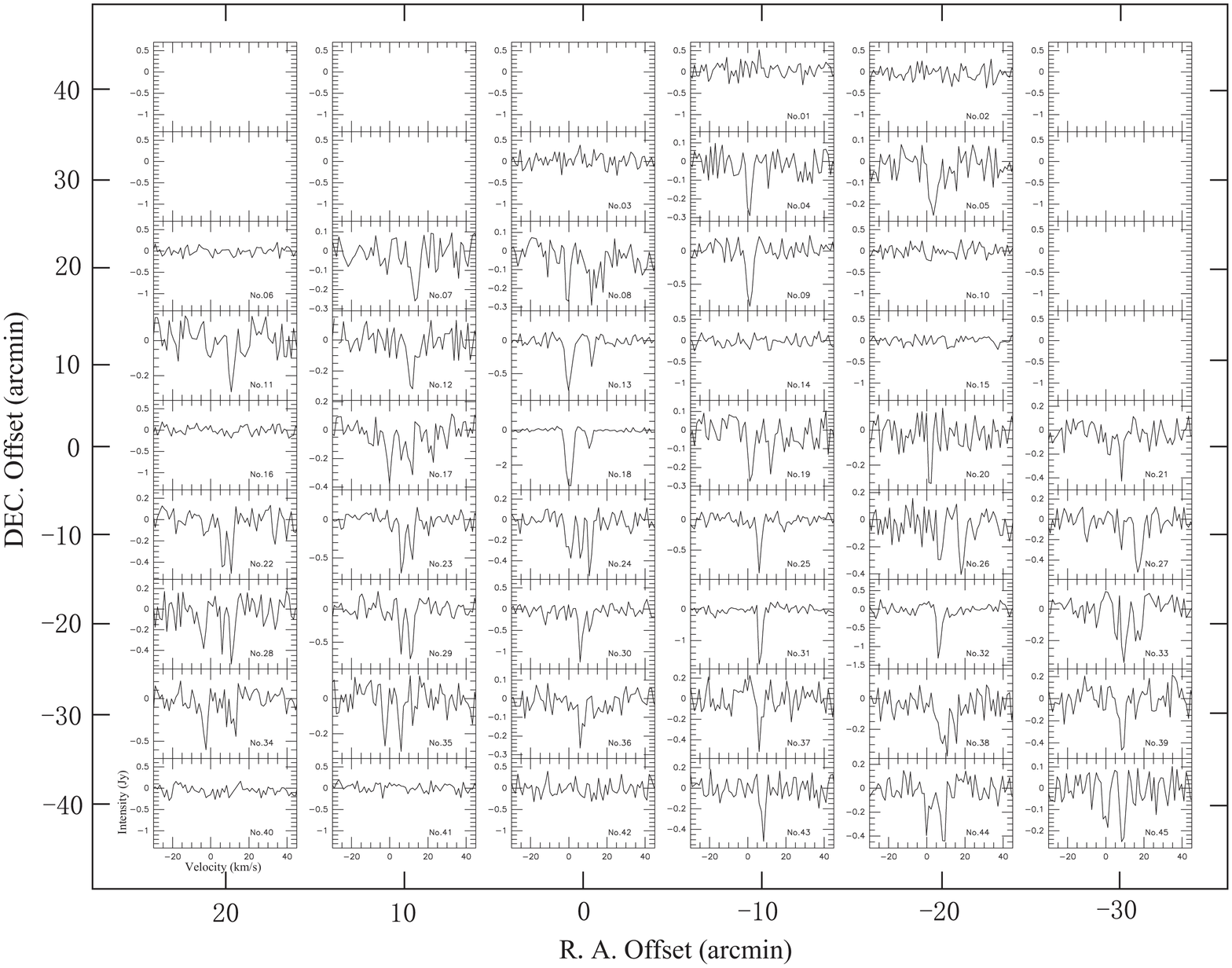}}
\hspace{1in} \subfigure[DR21/W75 GMC]{\label{dr21b}
\includegraphics[width=0.8\textwidth]{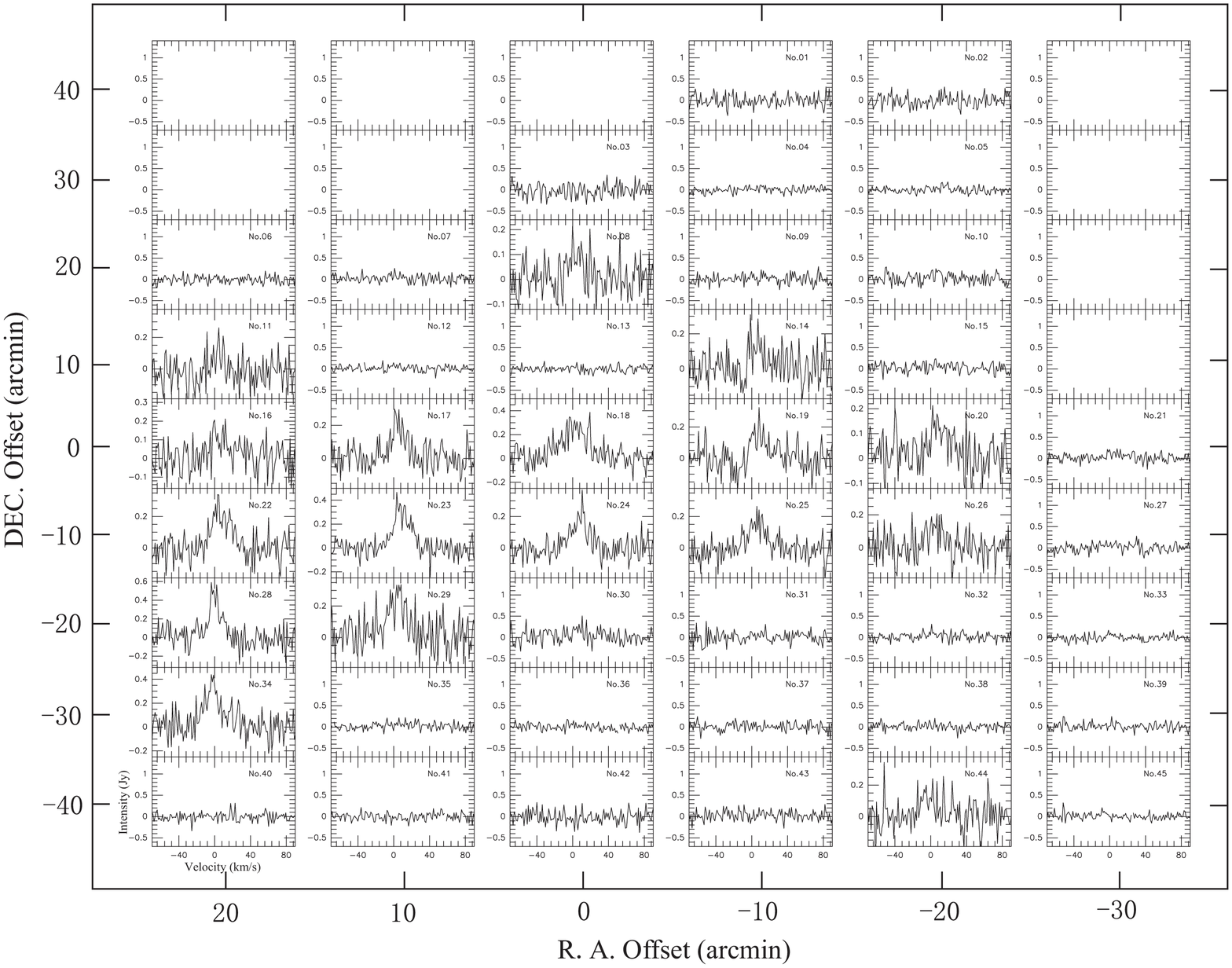}}
\caption{The spectral mosaic of (a) H$_{2}$CO absorption line and
(b) H110$\alpha$ RRL toward DR21/W75 GMC.}\label{dr21ab}
\end{figure}

\begin{figure}
\centering \subfigure[NGC2024/NGC2023 GMC]{\label{ngc2024a}
\includegraphics[width=0.65\textwidth]{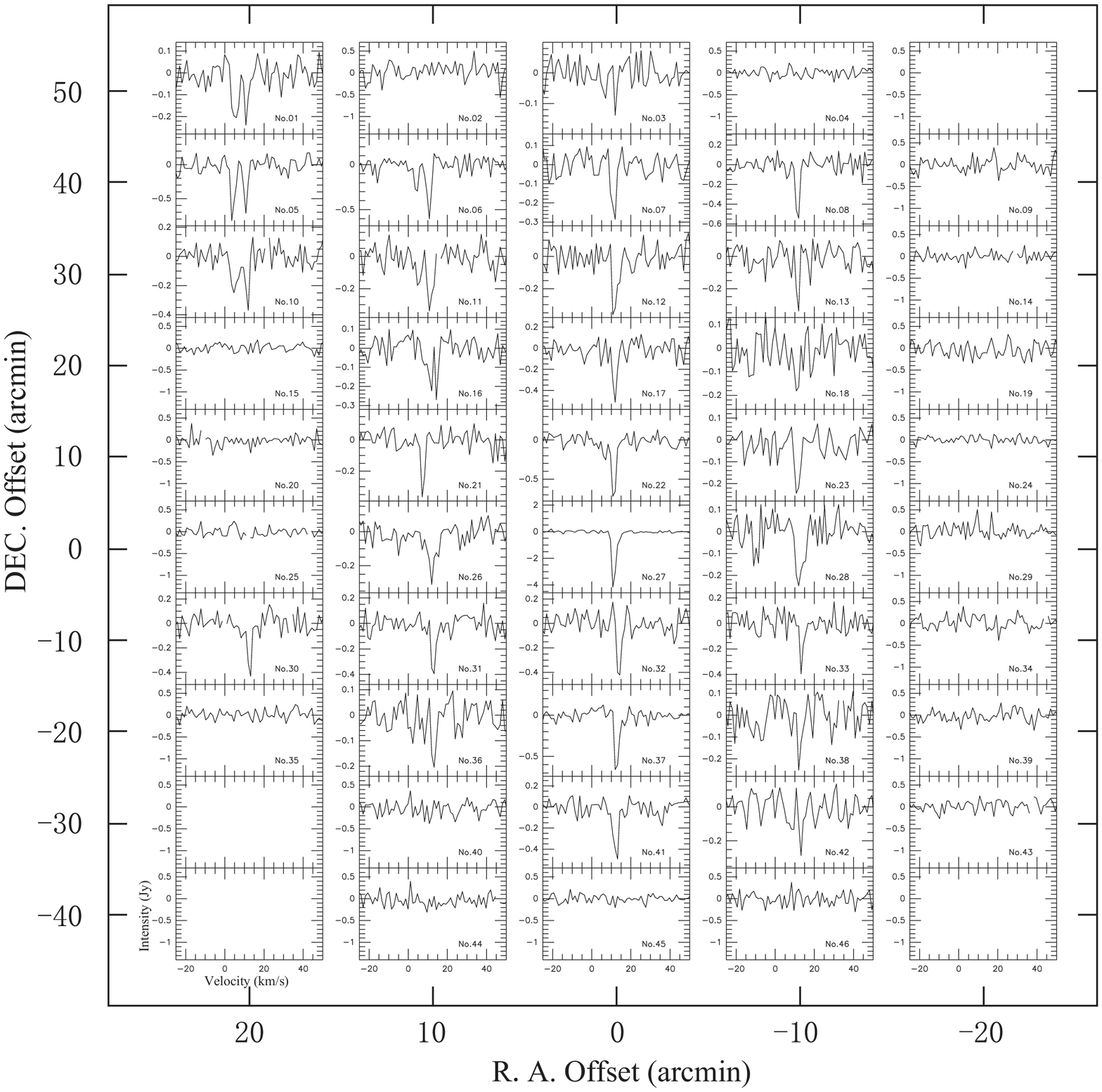}}
\hspace{1in} \subfigure[NGC2024/NGC2023 GMC]{\label{ngc2024b}
\includegraphics[width=0.65\textwidth]{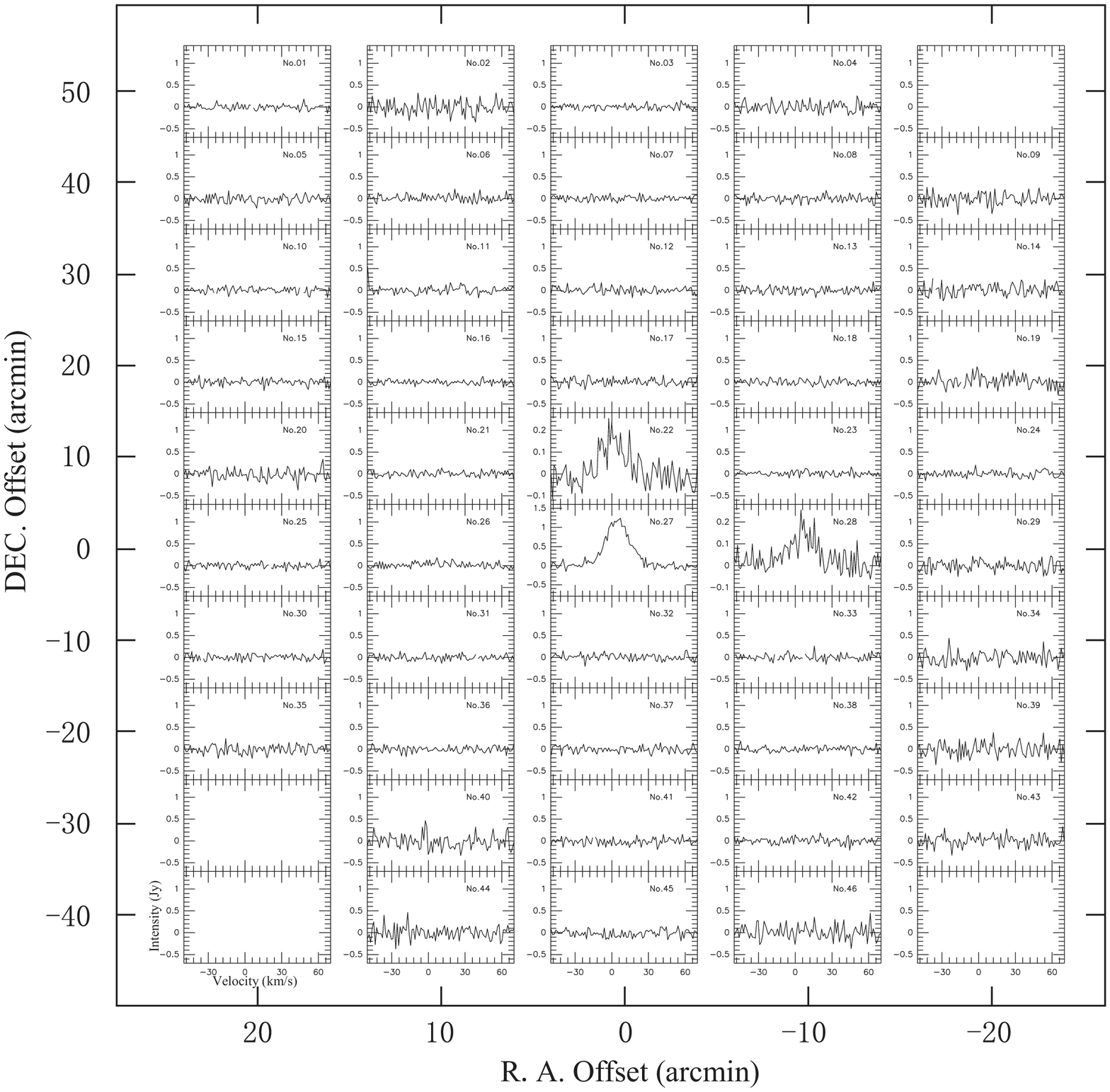}}
\caption{The spectral mosaic of (a) H$_{2}$CO absorption line and
(b) H110$\alpha$ RRL toward NGC2024/NGC2023 GMC.}\label{ngc2024ab}
\end{figure}

Table \ref{tbl-total} lists ID numbers corresponding to the (0, 0)
offset position in Figs. \ref{w49ab}, \ref{w3ab}, \ref{dr21ab} and
\ref{ngc2024ab} for the four sources. Columns 3 and 4 show
equatorial coordinates for the (0, 0) offset position. The distance
in Column 5 comes from references ($1$, $2$, $3$ and $4$) in Column
10. The size ($\alpha\times\delta$) in Column 6 is the approximate
survey region, which is much larger than in previous studies. Column
7 lists the $H_{2}$ clouds mass and Column 8 the total integration
time in the ON-position for every point source. In Column 9, $A$ is
the number of all observational positions toward the sources, $B$ is
the number of H$_{2}$CO absorption lines, and $C$ is the number of
H110$\alpha$ RRLs. Hence, the detection rate is 56.95\% for
H$_{2}$CO absorption lines and 10.60\% for H110$\alpha$ RRLs.

Figs. \ref{w49ab}, \ref{w3ab}, \ref{dr21ab} and \ref{ngc2024ab}
present spectral mosaics of H$_{2}$CO absorption lines and
H110$\alpha$ RRLs toward W49, W3, DR21 and NGC2024 GMCs. The
relative observational positions and the corresponding parameters
are listed in Tables \ref{tbl-w49}, \ref{tbl-w3}, \ref{tbl-dr21} and
\ref{tbl-ngc2024}. The offset position is indicated on the relative
coordinate axis and the step size is 10$'$. The velocity components
were identified by Gaussian fitting. All spectra are included,
regardless of whether they contain signals. There are many blank
panels for which sources could not be detected. Finally, all
H$_{2}$CO absorption lines exhibited are stronger than 3$\sigma$ in
intensity. However, some of the H110$\alpha$ RRLs did not reach
3$\sigma$ in intensity, as denoted by $a$ in Tables \ref{tbl-w49},
\ref{tbl-w3}, \ref{tbl-dr21} and \ref{tbl-ngc2024}. For faint ($<
3\sigma$) H110$\alpha$ RRLs, the spectra are shown for information
only and were not studied further.

Tables \ref{tbl-w49}, \ref{tbl-w3}, \ref{tbl-dr21} and
\ref{tbl-ngc2024} list the relevant parameters for H$_{2}$CO
absorption lines and H110$\alpha$ RRLs. The serial number and
coordinate offset are listed in Columns 1 and 2 for the
corresponding spectral mosaics (Figs. \ref{w49ab}, \ref{w3ab},
\ref{dr21ab} and \ref{ngc2024ab}). Columns 3--6 lists parameter data
for H$_{2}$CO absorption lines, while those for H110$\alpha$ RRLs
are in Columns 11--14. Columns 3 and 11 list line-of-sight velocity
data relative to the Sun. Columns 4 and 12 list the integration
intensity flux for each velocity component. Columns 5 and 13 list
line width (FWHM) data with $\Delta V$. Columns 6 and 14 list
spectral intensity data. Columns 7--10 list the CBT at 4.8 GHz, the
H$_{2}$CO optical depth, and the H$_{2}$CO and H$_{2}$ column
density, respectively. The optical depth ($\tau_{app}$) range is
approximately 0.007--0.188, so H$_{2}$CO is optically thin. The
column density ($N(H_{2}CO)$) range is approximately
$0.121\times10^{13}cm^{-2}-3.59 \times10^{13}$ $cm^{-2}$. In
addition, $N$ indicates that the corresponding spectra could not be
detected. H110$\alpha$ RRL data with a signal intensity of $<
3\sigma$ are denoted by $a$.

\begin{figure}[h!!!]
   \centering
   \includegraphics[width=6.5cm, angle=0]{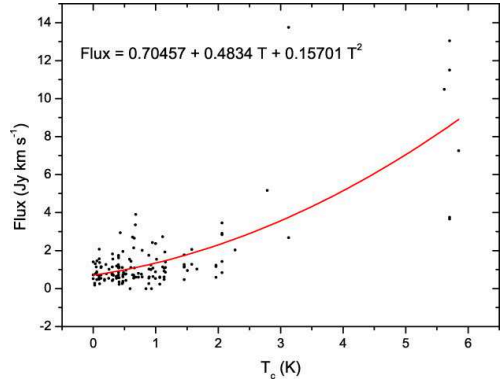}
   \caption{The relation between integration intensity of H$_{2}$CO and continuum brightness temperature
   of 4.85 GHz for W49, W3, DR21 and NGC2024 GMC.}
   \label{flux_t}
   \end{figure}

\begin{figure}[h!!!]
   \centering
   \includegraphics[width=6.5cm, angle=0]{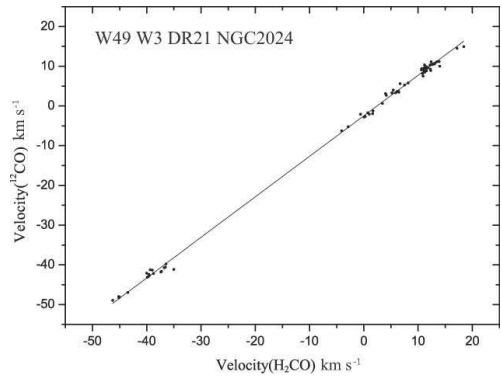}
   \caption{Correlation between velocity of H$_{2}$CO absorption line and $^{12}$CO
   emission line for W49, W3, DR21/W75 and NGC2024/NGC2023 GMCs. From the data line,
   we can find the line passes through (0, 0) point and the points almost
   distribute on or near the line. So the relation between them is distinct.}
   \label{velocity}
   \end{figure}

\begin{figure}[h!!!]
   \centering
   \includegraphics[width=6.5cm, angle=0]{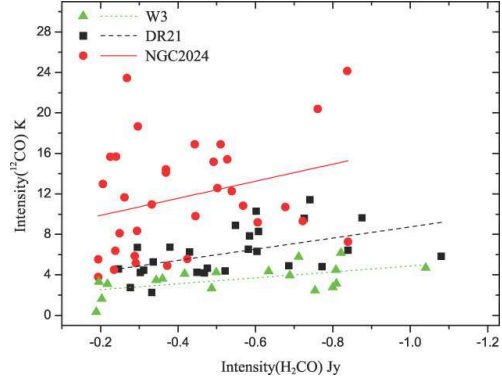}
   \caption{For W3, DR21/W75 and NGC2024/NGC2023 GMCs, correlation coefficient respectively being 0.558, 0.499 and
   0.297 between intensities of $^{12}$CO emission line and H$_{2}$CO absorption line.}
   \label{intensity}
   \end{figure}

\begin{figure}[h!!!]
   \centering
   \includegraphics[width=6.5cm, angle=0]{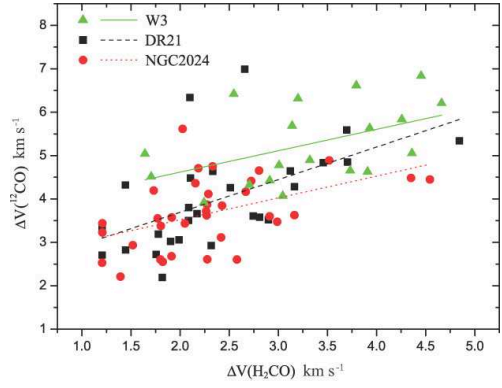}
   \caption{For W3, DR21/W75 and NGC2024/NGC2023 GMCs, correlation coefficient respectively being 0.480, 0.556
   and 0.478 between line widths of $^{12}$CO emission line and H$_{2}$CO absorption line.}
   \label{fwhm}
   \end{figure}

Fig. \ref{gmc} shows an overlay of the integration intensity for
H$_{2}$CO and $^{12}$CO contours onto the 8.28-$\mu$ m MSX color
map. And, several representative objects are indicated in the maps.
From Fig. \ref{gmc}, we find that the large area distributions of
them are consistent, but there are some off-peak discrepancy between
$^{12}$CO peaks and others. We also compare the relations between
velocity (Fig. \ref{velocity}), integration intensity (Fig.
\ref{flux}), intensity (Fig. \ref{intensity}) and line width (Fig.
\ref{fwhm}) to look for the relations between H$_{2}$CO and
$^{12}$CO. Moreover, we also overlay the H$_{2}$CO contours onto the
4.8 GHz continuum temperature contours in Fig. \ref{continuum}. From
Fig. \ref{continuum}, we find that there is a good consistent
morphologic distribution between H$_{2}$CO and CBT, and we also made
a relation (Fig. \ref{flux_t}) between integration intensity of H2CO
and CBT to derive a formula (Eqn. \ref{eq:flux}). Basing above, we
suggest that H$_{2}$CO contours are more strongly correlated with
the distribution of the continuum components than $^{12}$CO, so that
it is possible to produce the off-peak discrepancy. The character of
these four GMCs is analyzed in Section \ref{data-dscption-analy}.

\section{Results and discussion} \label{data-dscption-analy}

\subsection{GMC descriptions} \label{four}

\subsubsection{W49 GMC} \label{w49gmc}

The W49 GMC radio source was discovered in a 21-cm continuum survey
by \citet{wes58}. The W49 GMC complex consists of a thermal
component (W49A) and a nonthermal component (W49B). W49A is one of
the most luminous galactic giant radio HII regions
\citep{dre84,dep97}, and W49B is a supernova remnant (SNR). W49B and
W49A are separated by 12.5$'$ along an east--west line at a
kinematic distance of 11.4 kpc \citep{gwi92}. At this distance,
10$'$ = 33.16 pc. A giant-scale area (approx. 50$'$ $\times$ 50$'$)
was surveyed toward W49 GMC. During 966 minutes of integration, we
observed five H$_{2}$CO absorption lines and two H110$\alpha$ RRLs.
In our spectra (Fig.\ref{w49a}), three H$_{2}$CO components of
$\thicksim$15.9, $\thicksim$40.9 and $\thicksim$64.3 km s$^{-1}$
were detected. \citet{bro01} mapped the HII region of W49A and W49B
with 21-cm HI data, which revealed velocities of $\thicksim$4 and
$\thicksim$7 km s$^{-1}$ toward W49A, and $\thicksim$40 and
$\thicksim$60 km s$^{-1}$ toward W49A and W49B. H$_{2}$O emission
lines were also observed at $\thicksim$39 and $\thicksim$60 km
s$^{-1}$ in W49A \citep{buh69}. Furthermore, three absorption
features at velocities of $\thicksim$15, $\thicksim$40 and
$\thicksim$60 km s$^{-1}$ were found for 18-cm OH absorption lines
\citep{pas73}. It is possible that the W49 GMC has an intricate
kinematic structure or different velocity subclouds piled up in the
line of sight. The multiple velocity components may arise from the
Sagittarius spiral arm clouds \citep{bro01}. There is no clear
evidence that W49B is closer to the Sun than W49A. However, many
researchers are interested in whether W49A and W49B are physically
correlated. Considering the uniform velocity components
$\thicksim$4.5, $\thicksim$10.5, $\thicksim$15.9, $\thicksim$40.9
and $\thicksim$64.3 km s$^{-1}$ from W49A and W49B, a physical
association between them can be inferred.

\subsubsection{W3 GMC} \label{w3gmc}

W3 GMC, at 1.95 kpc from the Sun \citep{xu06}, lies to the western
edge of W4 GMC. At this distance, 10$'$ = 5.672 pc. W3 GMC is made
up of three knots of molecular gas known as W3 Main, W3 North, and
W3 OH. The Central Cluster, located between W3 Main and W3 OH,
contains a large number of Class II YSOs \citep{ruc07}. A
giant-scale area (approx. 70$'$ $\times$ 90$'$) was surveyed toward
W3 GMC. We detected 19 H$_{2}$CO absorption lines and four
H110$\alpha$ RRLs during 2370 minutes of integration. The W3 GMC is
a complex of massive star formation regions, where there are strong
H$_{2}$CO absorption lines, H110$\alpha$ RRLs, $^{12}$CO emission
lines, MSX sources and a CBT. We can clearly distinguish two cores
for W3 GMC. At the junction (W3(OH)) of the two clouds, we cannot
detect H$_{2}$CO absorption lines and the CBT is low, but the
integration intensity for $^{12}$CO clouds and MSX sources is
relatively strong. We hypothesize that many young stars in W3(OH)
are surrounded by a thin gas envelope. The velocity for H$_{2}$CO
and H110$\alpha$ ranges from approximately $-46.0$ to $-35.0$ km
s$^{-1}$, which is similar to HI observations \citep{rea81}. Table
\ref{tbl-w3} reveals a strange phenomenon: a sharp velocity gradient
is apparent, which is consistent with $J$ = 1-0 $^{12}$CO
observations \citep{thr85}. From northwest to southeast, the
velocity varies strongly from $-35.02$ km s$^{-1}$ (No. 08) to
$-45.14$ km s$^{-1}$ (No. 37) (Fig.~\ref{w3a} and Table
\ref{tbl-w3}).

\subsubsection{DR21/W75 GMC} \label{dr21gmc}

DR21/W75 GMC is located in the Cygnus constellation, approximately
3.0 kpc from the Sun \citep{cam82}. At this distance, 10$'$ = 8.727
pc. The MSX color map reveals that DR21/W75 GMC exhibits a complex
and dispersive structure, with many separate subclusters assembled
together. We surveyed DR21 and W75 GMCs, which include W75N, W75,
DR21(OH), DR21, L906E and Diamond Ring (source name). DR21 and W75
GMCs are associated with massive dense cores and are separated by
approximately 30$'$ \citep{wil90,shi03}. A giant-scale area (approx.
60$'$ $\times$ 90$'$) was surveyed toward DR21/W75 GMC. We detected
34 H$_{2}$CO absorption lines and 8 H110$\alpha$ RRLs during 2742
minutes of integration. The velocity components are multiple and
rather intricate (Fig.~\ref{dr21a}). The velocity is smaller for the
northeastern and larger for the southwestern section than for the
central section. A velocity gradient exists within these subclouds,
and probably arises from GMC rotation. A UCHII region was detected
using H110$\alpha$ RRLs as a tracer. The data in Table
\ref{tbl-dr21} reveal that the H110$\alpha$ RRL intensity is so weak
(indicated by $a$) that we could not obtain a good signal, even with
long-time integration. In addition, we only detected part of the
DR21/W75 GMC. In particular, for the western part there is a giant
and strong MSX region, while the $^{12}$CO cloud is relatively
faint. To determine whether or not there is association between
H$_{2}$CO and $^{12}$CO clouds and MSX sources, the western edge of
the DR21/W75 GMC should be observed.

\subsubsection{NGC2024/NGC2023 GMC} \label{ngc2024gmc}

NGC2024/NGC2023 GMC is situated in Orion B at a distance of 415 pc
\citep{men07}. At this distance, 10$'$ = 1.207 pc. The
NGC2024/NGC2023 GMC is a bright emission nebula crossed by a
prominent dust lane. NGC2024 GMC includes a number of protostars
along the star-forming ridge extending in a north--south direction
coincident with an HII region, which is in front of filamentary
shaped dense molecular material \citep{gau92}. A giant-scale area
(approx. 50$'$ $\times$ 100$'$) was surveyed toward NGC2024/NGC2023
GMC. The area has similar velocity components and integration
intensity contours to those reported by \citet{coh83}, but our
observational instrument was more sensitive than theirs. We detected
28 H$_{2}$CO absorption lines and two H110$\alpha$ RRLs during 3378
minutes of integration. The H$_{2}$CO velocity is approximately
11.70 km s$^{-1}$ and comprises a single component (Table
\ref{tbl-ngc2024}). To the north of NGC2024/NGC2023, no strong
infrared background can be detected and both the NRAO/VLA Sky Survey
radio contour plot and the CBT at 4.8 GHz are much fainter than in
the central section, whereas H$_{2}$CO absorption and $^{12}$CO
emission lines are widely distributed in this area. Hence, it is
possible that the northern H$_{2}$CO absorption lines arise from CMB
excitation.

\subsection{Comparison of H$_{2}$CO distributions and the CBT} \label{distribution}

It is well known that the H$_{2}$CO absorption is strongly biased by
the high CBT which is being absorbed by the H$_{2}$CO molecules
along the line of sight to the source of the continuum. Here we will
give the empirical relationship between H$_{2}$CO and CBT. In Fig.
\ref{continuum}, continuum contours are overlaid on the H$_{2}$CO
contour maps for W49, W3, DR21 and NGC2024 GMCs. The morphology of
the H$_{2}$CO and continuum distributions matches very well. The
H$_{2}$CO peaks are biased to the CBT peaks. For the offset(0, 0)
positions of four GMCs, the CBTs are so high that H$_{2}$CO maybe
mainly come from the CBT collision excitation. According to Tables
\ref{tbl-w49}, \ref{tbl-w3}, \ref{tbl-dr21} and \ref{tbl-ngc2024},
the H$_{2}$CO column density shows an irregular distribution that
does not match the morphology of H$_{2}$CO intensity contours. By
comparing the integration intensity contours for H$_{2}$CO with the
CBT for GMCs (Fig. \ref{flux_t}) and by polynomial fitting, we
obtained the following equation:
\begin{eqnarray}
    \label{eq:flux}
    Flux(H_{2}CO) = 0.70457 + 0.4834 T_{c} + 0.15701 T_{c}^{2},
\end{eqnarray}
where $Flux(H_{2}CO)$ is integration intensity of H$_{2}$CO, and
$T_{c}$ is 6-cm CBT. This further suggests in quantity that the
H$_{2}$CO intensity is strongly influenced by the CBT background and
that the contribution of CMB excitation to the intensity is very
weak.

\subsection{Comparison of H$_{2}$CO, $^{12}$CO and MSX data} \label{comparison}

The velocity correlation between H$_{2}$CO and $^{12}$CO is
described in Fig. \ref{velocity} for W49 , W3, DR21 and NGC2024
GMCs. The fitting line passes through (0, 0) and the points are
distributed almost on or near the line.

\begin{figure}[h!!!]
   \centering
   \includegraphics[width=6.5cm, angle=0]{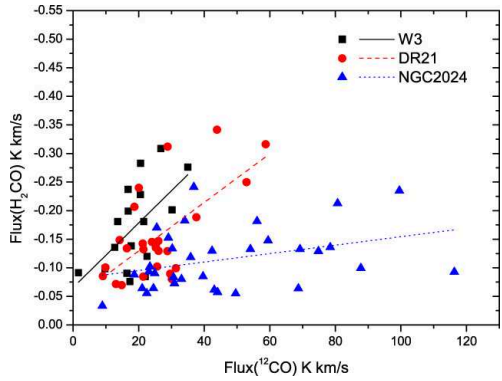}
   \caption{For W3, DR21/W75 and NGC2024/NGC2023 GMCs, the correlation between the H$_{2}$CO integration intensity and
   the $^{12}$CO integration intensity.}
   \label{flux}
   \end{figure}

The correlations of integration intensity between H$_{2}$CO and
$^{12}$CO are plotted in Fig. \ref{flux} (ignoring these points of
$|Flux(H_{2}CO)|>5 Jy km s^{-1}$) for W3, DR21 and NGC2024 GMCs. The
best least-squares fit to a straight line for W3, DR21 and NGC2024
GMCs GMC data sets yield respectively
\begin{eqnarray}
\begin{split}
    \label{eq:w3}
    W3: Flux(H_{2}CO) = 5.64\times10^{-3}Flux(^{12}CO)\\
    + 65.59\times10^{-3} K km s^{-1},
\end{split}
\end{eqnarray}
\begin{eqnarray}
\begin{split}
    \label{eq:dr21}
    DR21: Flux(H_{2}CO) = 4.25\times10^{-3}Flux(^{12}CO)\\
    + 44.36\times10^{-3} K km s^{-1},
\end{split}
\end{eqnarray}
\begin{eqnarray}
\begin{split}
    \label{eq:ngc2024}
    NGC2024: Flux(H_{2}CO) = 0.74\times10^{-3}Flux(^{12}CO)\\
    + 80.48\times10^{-3} K km s^{-1}.
\end{split}
\end{eqnarray}
Fig. \ref{flux} shows that the integration intensity relation
between H$_{2}$CO and $^{12}$CO for three GMCs is scattered. And the
equation coefficients are different but similar between
\citet{rod07} and our results. It is suggested that there exists
different physical condition for different GMCs, however, the
correlation between H$_{2}$CO and $^{12}$CO is inherent.

Fig.~\ref{intensity} shows the intensity correlation for H$_{2}$CO
and $^{12}$CO (ignoring these points of $|Intensity(H_{2}CO)|> 1.2
Jy$). For W3, DR21 and NGC2024 GMCs, the correlation coefficient is
0.558, 0.499 and 0.297, respectively. Thus, the linear relation is
better for W3 and DR21 GMCs than for NGC2024. The reason may be that
H$_{2}$CO absorption to the north of the NGC2024 GMC arises from CMB
excitation, whereas bright continuum sources are responsible for
H$_{2}$CO absorption in the other regions. The two excitation
mechanisms possibly produce different H$_{2}$CO intensities. The
intensity of H$_{2}$CO absorption is related to the molecular
density and background continuum sources. Using Origin Software to
draw and calculate, the Pearson's correlation coefficients for line
width between $^{12}$CO emission and H$_{2}$CO absorption are 0.480,
0.556 and 0.478 for W3, DR21 and NGC2024 GMCs respectively
(Fig.~\ref{fwhm}). The line width is greater for $^{12}$CO emission
than for H$_{2}$CO absorption on the whole. In general, there is
good correlation between $^{12}$CO and H$_{2}$CO clouds basing on
these correlation coefficients.

Comparison of H$_{2}$CO and $^{12}$CO contours and MSX data reveals
great uniformity in morphology in the large area, especially between
H$_{2}$CO and MSX data. The integrated intensity peaks are located
at almost the same positions, but the $^{12}$CO peaks are offset
from the H$_{2}$CO peaks. MSX clouds are always located within star
formation regions, where the hot core is the energy source for star
formation. Basing above, under ambient conditions, H$_{2}$CO and
$^{12}$CO clouds form a blanket around star formation regions in
morphology. This leads to a uniform structure that can be used as a
tracer for star formation regions.

\subsection{$^{12}$CO peak offset from H$_{2}$CO and CBT} \label{offset}

\begin{figure}
\centering \subfigure[W49 GMC]{\label{w494}
\includegraphics[width=0.39\textwidth]{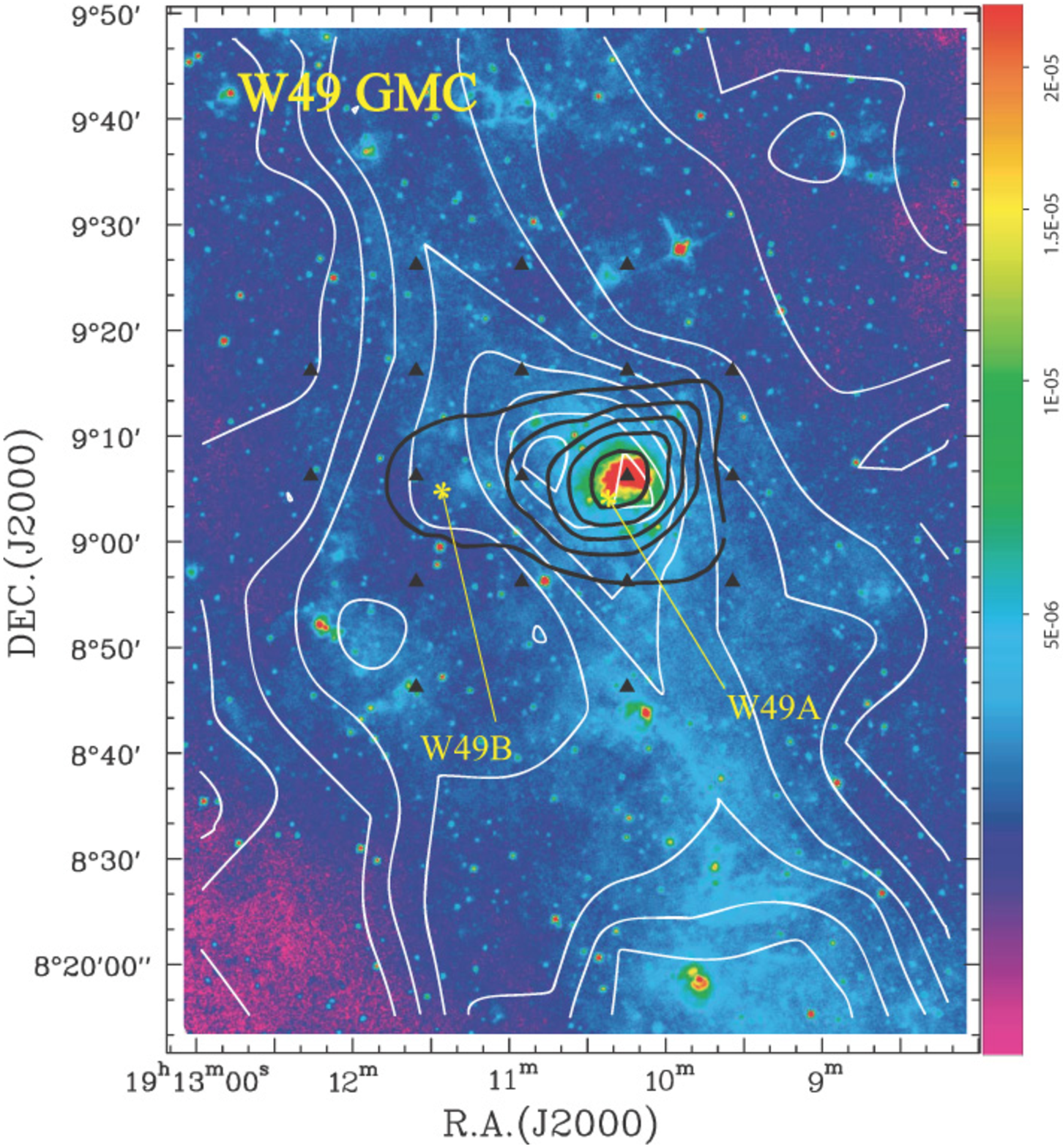}}
\hspace{1in} \subfigure[W3 GMC]{\label{w3}
\includegraphics[width=0.40\textwidth]{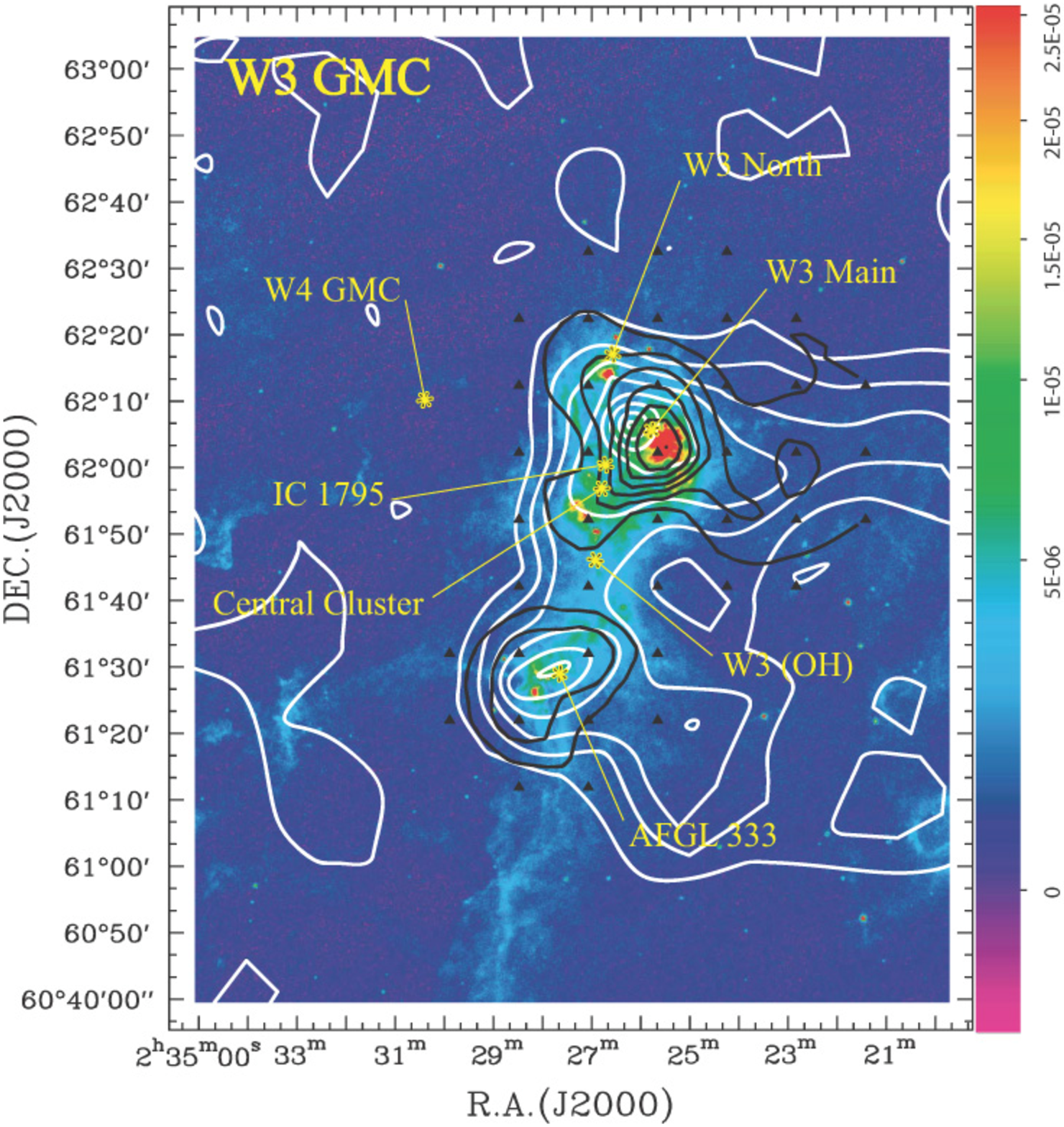}}
\hspace{1in} \raggedright \subfigure[DR21/W75 GMC]{\label{dr21}
\includegraphics[width=0.39\textwidth]{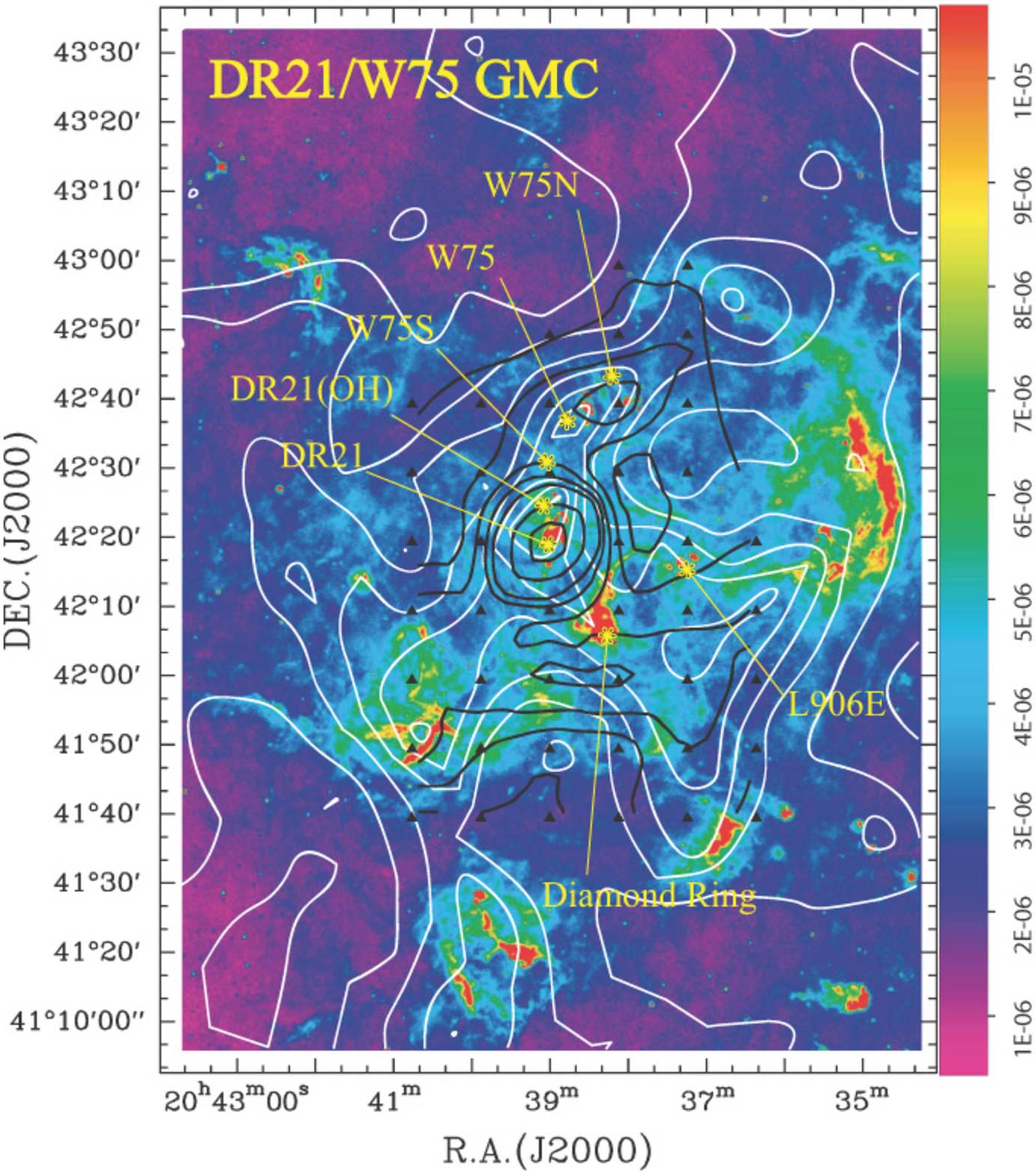}}
\hspace{1in} \subfigure[NGC2024/NGC2023 GMC]{\label{ngc2024}
\includegraphics[width=0.42\textwidth]{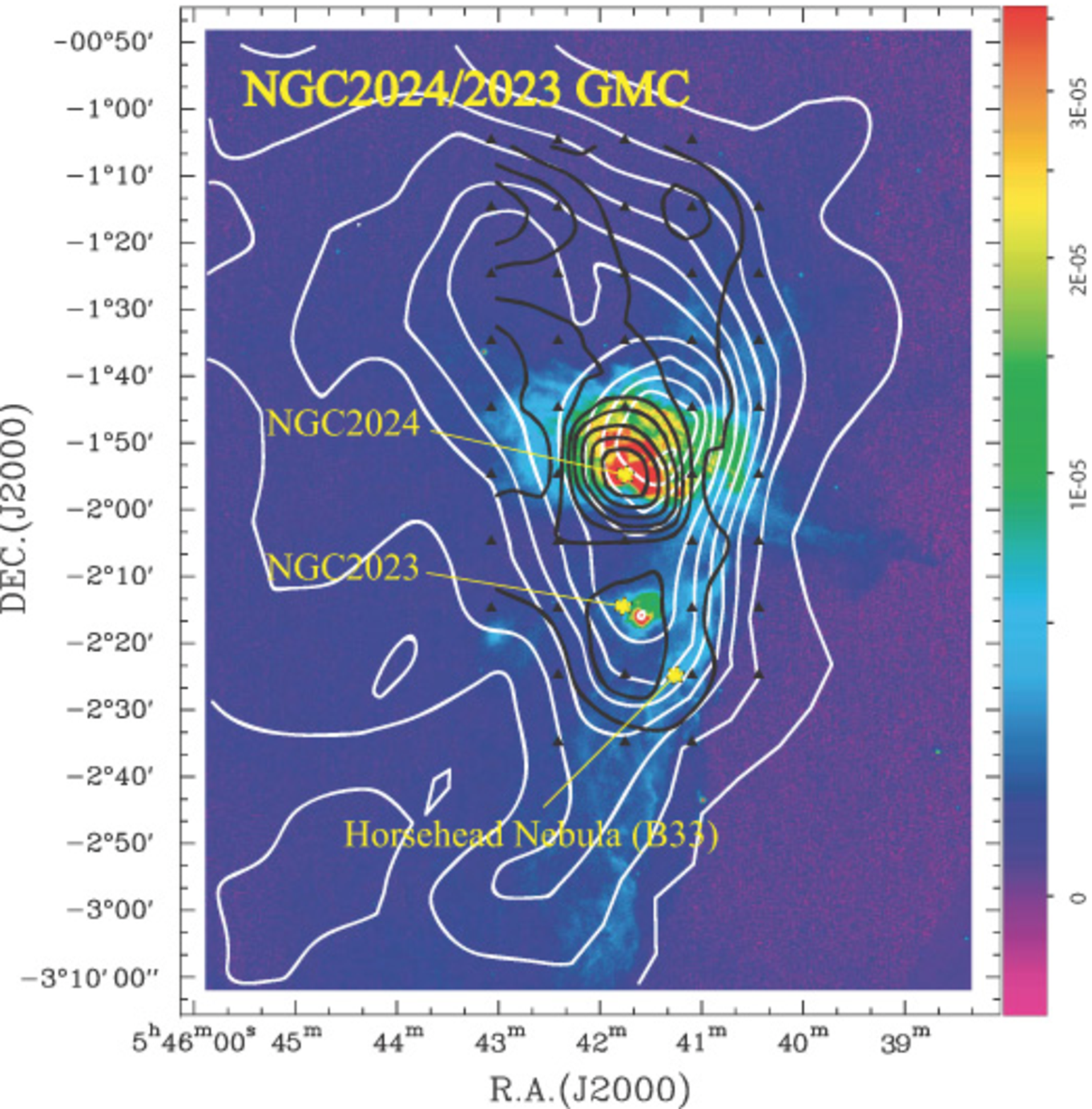}}
\caption{Contours and color-scale maps of integrated area toward
W49, W3, DR21/W75 and NGC2024/NGC2023. The black contours, the white
contours and the color-scale map respectively indicates the
integrated flux intensity of the H$_{2}$CO absorption line,
$^{12}$CO emission line and the mid-infrared 8.28-$\mu$m MSX source.
Triangle symbols indicate the location what we surveyed and the
coordinate of the offset (0, 0) position. (a), (0, 0): R.A. =
19$^{h}$10$^{m}$15$^{s}$.25, DEC. = 09$^\circ$06$'$08$''$.4. Levels
for black contours (H$_{2}$CO) with beam size 10$'$ are 29.15,
22.67, 16.19, 9.71 and 3.24 Jy km s$^{-1}$ from inside to outside,
while 100 to 20 by -10 K km s$^{-1}$ for white contours ($^{12}$CO)
with beam size 8$'$. The corresponding velocity component is from
-29.8 km s$^{-1}$ to 89.8 km s$^{-1}$ for H$_{2}$CO and $^{12}$CO.
(b), (0, 0): R.A. = 02$^{h}$25$^{m}$38$^{s}$.79, DEC. =
62$^\circ$02$^{'}$22$^{''}$.0. Levels for black contours (H$_{2}$CO)
with beam size 10$'$ are 6.34, 5.22, 4.11, 2.99, 1.87 and 0.75 Jy km
s$^{-1}$ from inside to outside, while 48 to 6 by -6 K km s$^{-1}$
for white contours ($^{12}$CO) with beam size 8$'$. The
corresponding velocity component is from -59.15 km s$^{-1}$ to
-20.15 km s$^{-1}$ for H$_{2}$CO and $^{12}$CO. (c), (0, 0): R.A. =
20$^{h}$39$^{m}$01$^{s}$.23, DEC. = 42$^\circ$19$^{'}$33$^{''}$.9.
Levels for black contours (H$_{2}$CO) with beam size 10$'$ are
14.92, 10.00, 5.08, 3.85, 2.62, 1.39 and 0.16 Jy km s$^{-1}$ from
inside to outside, while 81 to 4 by -11 K km s$^{-1}$ for white
contours ($^{12}$CO) with beam size 8$'$. The corresponding velocity
component is from -15.4 km s$^{-1}$ to 25.6 km s$^{-1}$ for
H$_{2}$CO and $^{12}$CO. (d), (0, 0): R.A. =
05$^{h}$41$^{m}$45$^{s}$.49, DEC. = -01$^\circ$54$^{'}$46$^{''}$.8.
Levels for black contours (H$_{2}$CO) with beam size 10$'$ are 8.69,
6.59, 4.50, 3.66, 2.83, 1.99, 1.15 and 0.31 Jy km s$^{-1}$ from
inside to outside, while 124 to 4 by -12 K km s$^{-1}$ for white
contours ($^{12}$CO) with beam size 8$'$. The corresponding velocity
component is from 0.33 km s$^{-1}$ to 20.48 km s$^{-1}$ for
H$_{2}$CO and $^{12}$CO.}\label{gmc}

\end{figure}

\begin{figure}
\centering \subfigure[W49 GMC]{\label{w49c}
\includegraphics[width=0.4\textwidth]{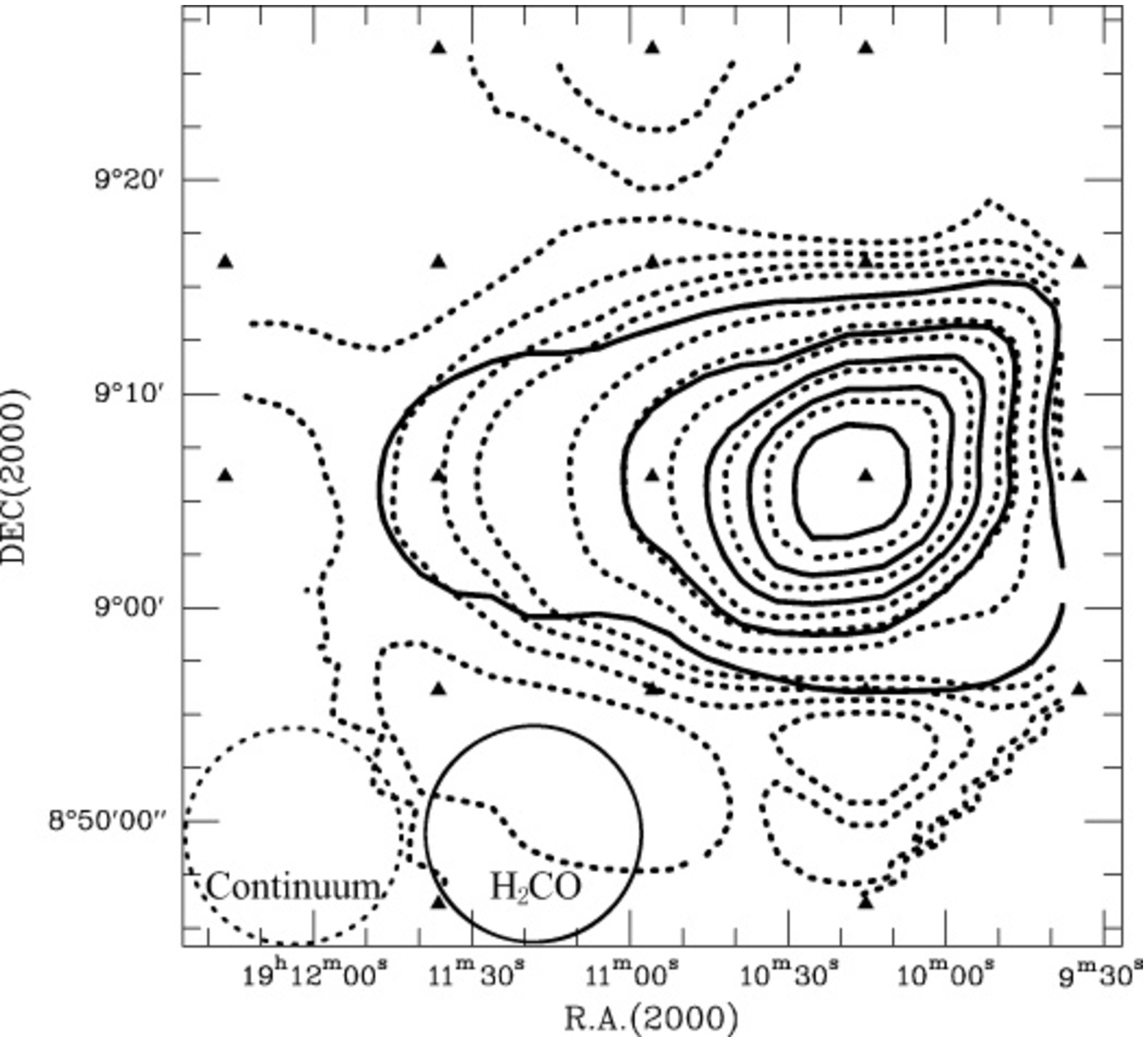}}
\hspace{1in} \subfigure[W3 GMC]{\label{w3c}
\includegraphics[width=0.35\textwidth]{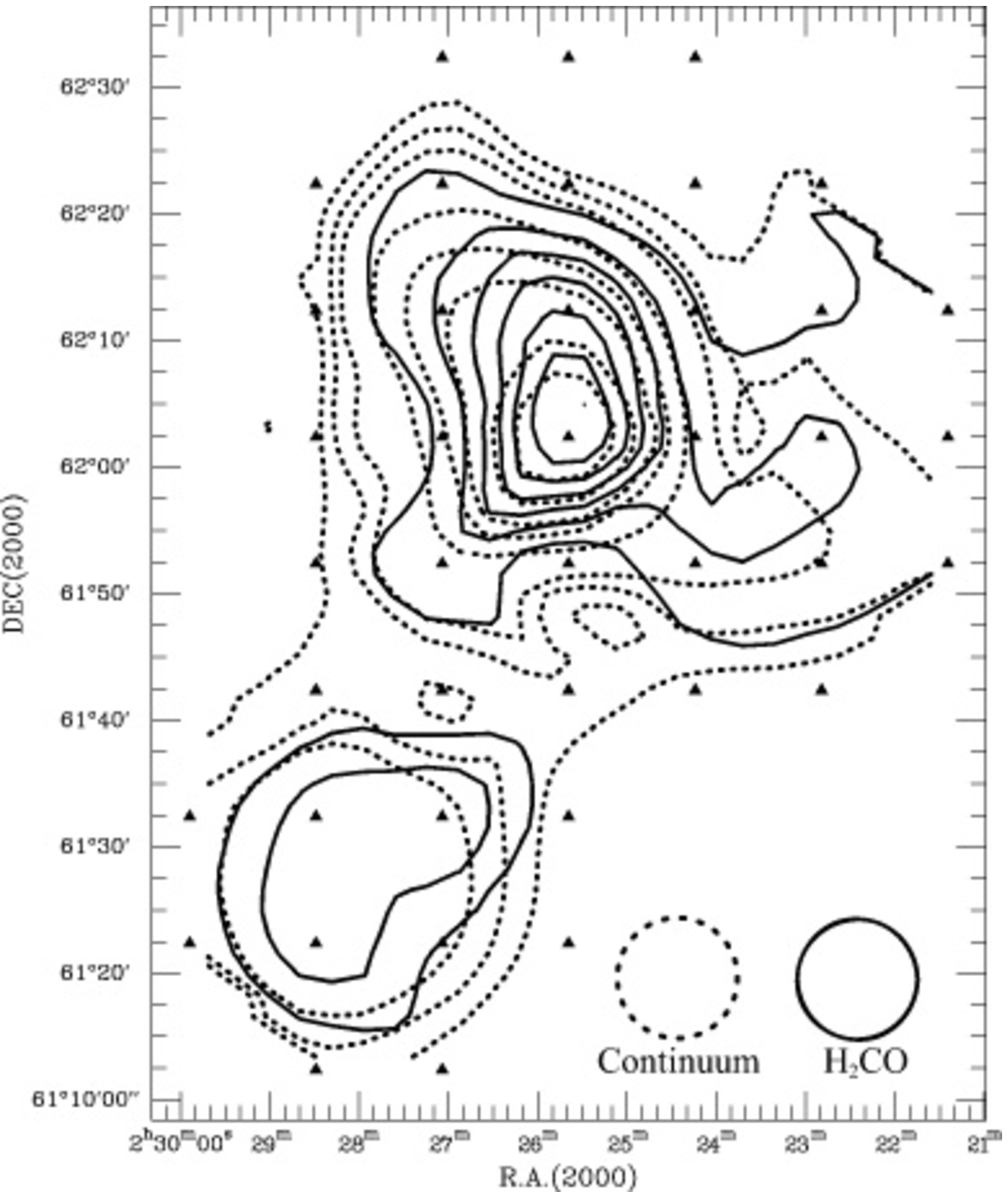}}
\hspace{1in} \raggedright \subfigure[DR21/W75 GMC]{\label{dr21c}
\includegraphics[width=0.4\textwidth]{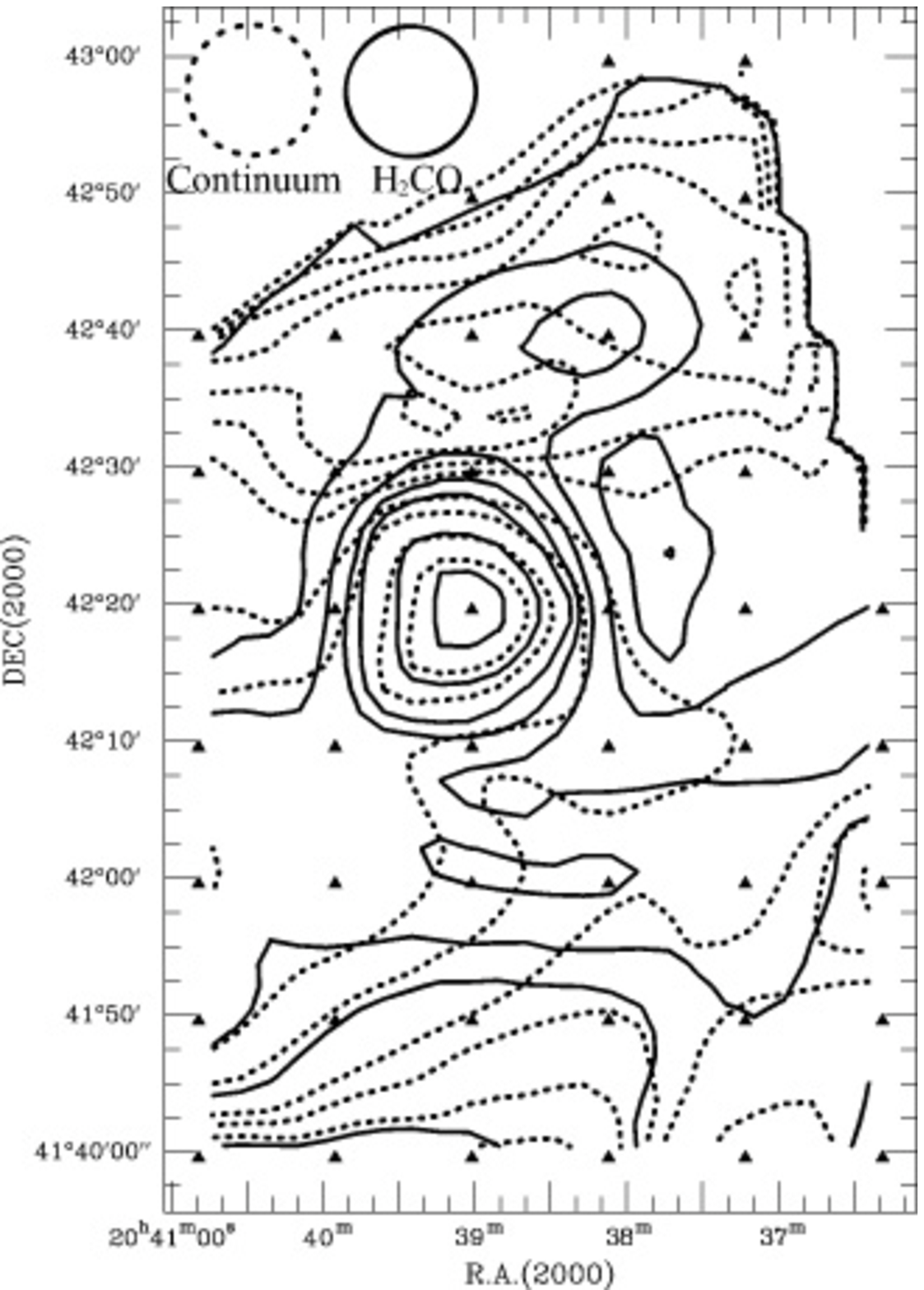}}
\hspace{1in} \subfigure[NGC2024/NGC2023 GMC]{\label{ngc2024c}
\includegraphics[width=0.31\textwidth]{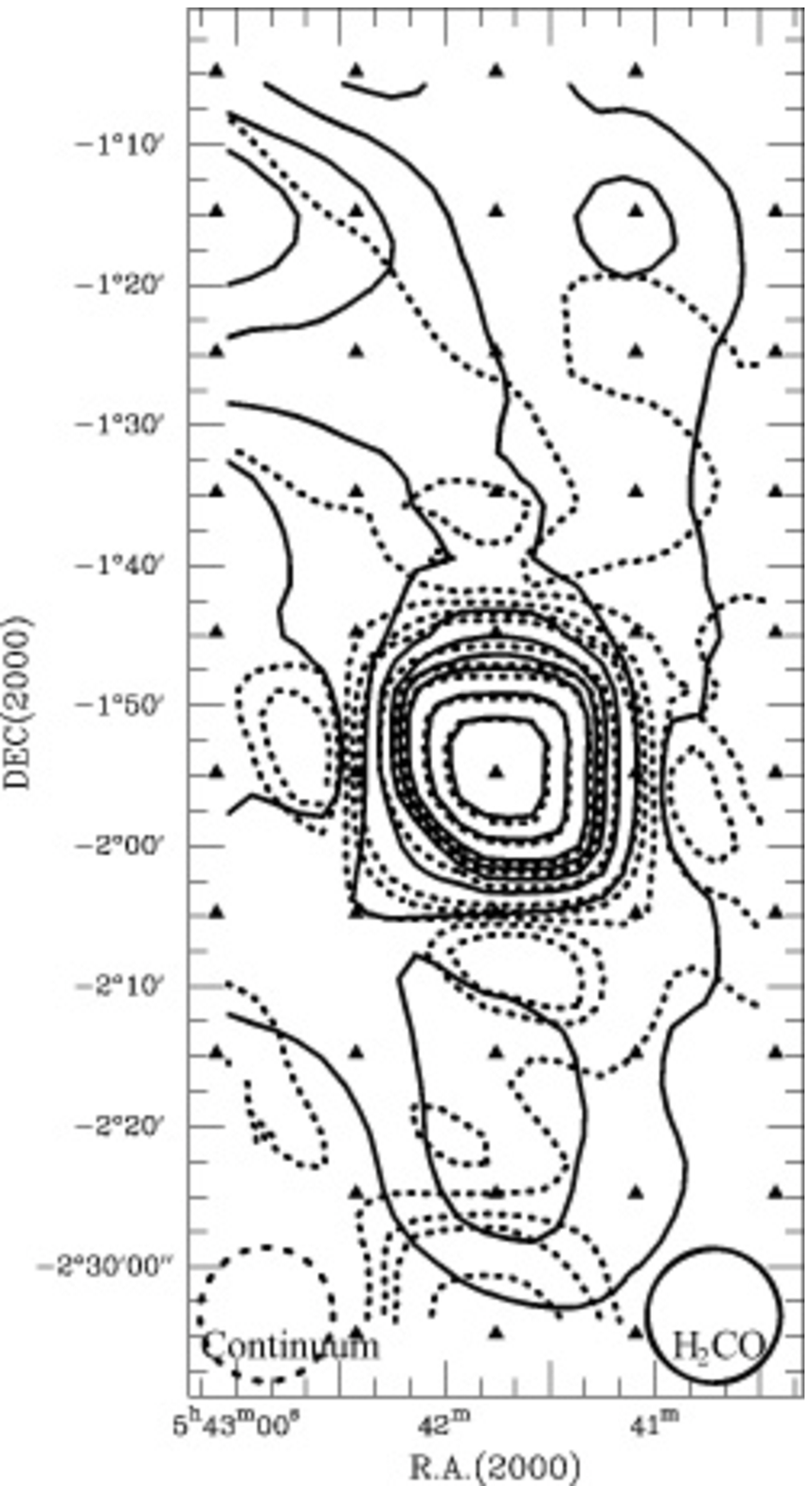}}
\caption{Integration intensity of H$_{2}$CO (Solid line) VS.
Continuum brightness temperature of 4.85 GHz (Dotted line) for W49
GMC. For continuum brightness temperature, the contour levels from
inside to outside are respectively (a) 4.601, 3.451, 2.301, 1.726,
1.150, 0.575, 0.431, 0.288 and 0.144 K, (b) 4.705, 3.529, 2.353,
1.765, 1.176, 0.588, 0.412 and 0.235 K, (c) 2.512, 1.884, 1.256,
0.942, 0.628, 0.534, 0.440, 0.345, 0.251, 0.157 and 0.063 K, and (d)
4.482, 3.361, 2.241, 1.681, 1.120, 0.560, 0.437, 0.314, 0.190 and
0.067 K, while they are same as Fig. \ref{gmc} for integration
intensity of H$_{2}$CO.}\label{continuum}
\end{figure}

For the large-area distribution, $^{12}$CO and H$_{2}$CO contours
are fairly similar in morphology, but there is a peak offset (about
10$'$ in Fig. \ref{w3} and \ref{ngc2024}) for the small-area
distribution. \citet{hei87} did not find a morphologic correlation
between $^{12}$CO and H$_{2}$CO data using a 100-m telescope, but
\citet{rod07} reported a discrepancy between $^{12}$CO and H$_{2}$CO
peaks toward the L1204 dust cloud using the Onsala 25-m telescope.
Their H$_{2}$CO clouds were against the CMB, but our sources are
against high CBT sources in HII regions. The background continuum
temperature has a strong effect on the H$_{2}$CO distribution as we
showed in Fig. \ref{continuum}, which reveals a fairly consistent
morphology between H$_{2}$CO and the continuum. However, Fig.
\ref{gmc} shows that $^{12}$CO peaks are offset from H$_{2}$CO
peaks, although this offset is different from that observed by
\citet{rod07}. Their H$_{2}$CO peaks were offset from the bright
continuum source, while our $^{12}$CO peaks are offset from the
bright continuum source. \citet{rod07} argued that the H$_{2}$CO
peak offset arises from photodissociation in the UV interstellar
radiation field. In our opinion, several factors explain the peak
offset. First, the H$_{2}$CO distribution is strongly biased by the
background CBT, while the strong HII region background has a
relatively weak impact on $^{12}$CO emission. Second, differences in
star formation regions and evolution stages between sources will
lead to discrepancy. Third, it is likely that H$_{2}$CO absorption
is optically thin ($<$ 0.188) and $^{12}$CO emission is optically
thick \citep{buc10}, so $^{12}$CO is a poor tracer for density.
Finally, the different resolution may lead to discrepancy.

\subsection{Statistical relationship between line width and intensity} \label{relation}

\begin{figure}[h!!!]
   \centering
   \includegraphics[width=6.5cm, angle=0]{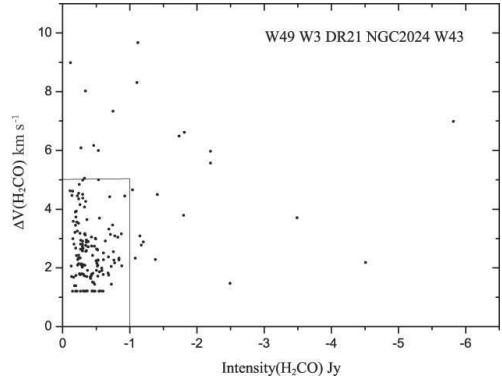}
   \caption{Correlation between line width and intensity of H$_{2}$CO absorption line at
   the each spectrum for W49, W3, DR21/W75 and NGC2024/NGC2023 GMCs£¬ and W43 GMC \citet{wu11}. Each point comes from every velocity
   component of spectrum. The points in the range
   from $-1.0$ to 0 Jy and the $\Delta V$ range from 1.206 to 5 km
   s$^{-1}$ hold 85.21\% of all points.}
   \label{fwhmintensity}
   \end{figure}

Fig.~\ref{fwhmintensity} shows a statistical diagram of the
relationship between line width and intensity for H$_{2}$CO
absorption lines toward W49, W3, DR21/W75 and NGC2024/NGC2023 GMCs,
and for W43 GMC observations made by \citet{wu11} using the same
dish. These data points are corresponding to 169 velocity components
from our detected 86 H$_{2}$CO absorption lines. The statistical
results show that the most (approximately 85.21\%) of the velocity
components of the H$_{2}$CO absorption lines are in the intensity
range from $-1.0$ to 0 Jy and in the $\Delta V$ range 1.206--5.0 km
s$^{-1}$.

In Fig.~\ref{fwhmintensity} the width for most H$_{2}$CO absorption
lines is relatively narrow and may be narrower than the 1.206 km
s$^{-1}$ for the velocity channel. However, a few lines are rather
wide, even up to 9.67 km s$^{-1}$. According to \citet{bie82}, the
width of all H$_{2}$CO absorption lines observed in this paper far
exceeds the thermal line width for any reasonable gas kinetic
temperature (e.g. $\Delta V$ (thermal) = 0.3 km s$^{-1}$ for
H$_{2}$CO at 30 K), so the thermal broadening mechanism can be
ignored. We suggest that line broadening is mainly the result of
turbulence and velocity dispersion along any given line of sight.
Thus, H$_{2}$CO line broadening should mainly reflect the blending
of many velocity components.

\section{Summary} \label{summa}

H$_{2}$CO absorption lines are important tracers for detecting the
ambient conditions in star formation regions. So far very few people
have carried out such large-scale H$_{2}$CO mapping as ours,
especially for GMCs. We conducted discontinuous observations toward
four GMCs from October 2009 to January 2011 using the Nanshan 25-m
telescope. Long-time integration observations and analysis revealed
the following results.

For W49, W3, DR21/W75 and NGC2024/NGC2023 GMCs, we observed 151
points using a beam width of 10$'$ and integration for 9456 min, and
found 86 H$_{2}$CO absorption lines and 16 H110$\alpha$ RRLs. We
processed H$_{2}$CO absorption lines for four large-area mappings.

We described and gave some relevant physical parameters which are
respectively flux, velocity, line width, intensity, CBT, apparent
depth and column density for H$_{2}$CO absorption lines,
H110$\alpha$ RRLs and continuum and clump's $H_{2}$-mass for GMCs.
Some good correlation coefficients between H$_{2}$CO and $^{12}$CO
were gained in terms of velocity components, line width and
intensity.

In the large area, comparisons among H$_{2}$CO and $^{12}$CO
contours, the 8.28-$\mu$ m MSX color map and CBT at 4.8 GHz revealed
a consistent distribution. Regions with a high CBT had much higher
collision excitation rates for H$_{2}$CO. However, in the small
area, H$_{2}$CO and $^{12}$CO peaks were not located at the same
position. It is likely that the H$_{2}$CO distribution is strongly
biased by the background CBT, while the strong HII region of the
background has a relatively weak impact on $^{12}$CO emission.

Many other results were observed for these four GMCs. E.g., W49B is
a nonthermal SNR and has nearly the same velocity components as the
thermal W49A. At the junction of two clouds for W3 GMC, we could not
detect H$_{2}$CO absorption lines, whereas the $^{12}$CO and MSX
data were relatively strong. DR21/W75 GMCs had a velocity gradient,
possibly arising from GMC rotation. It is possible that H$_{2}$CO
absorption lines to the north of NGC2024/NGC2023 GMC arise from CMB
excitation.

A statistical diagram of the relation between line width and
intensity was constructed for H$_{2}$CO absorption lines.
Approximately 85.21\% of the velocity components of H$_{2}$CO
absorption lines were distributed in the intensity range from $-1.0$
to 0 Jy and in the $\Delta V$ range 1.206--5.0 km s$^{-1}$.

\begin{acknowledgements} \footnotesize{We would like to thank all
staff of Nanshan Observatory for observations throughout day and
night. We are grateful for X. W. Zheng's assistance with providing
useful insight into several aspects of this work. And we are
grateful for T. M. Dame for providing and processing necessary
$^{12}$CO data. This work is supported by the National Natural
Science Foundation of China (Grant Nos. 10778703 and 10873025) and
the Program of the Light in China Western Region (LCWR; Nos.
RCPY200605 and RCPY200706). MSX research has made use of the NASA/
IPAC Infrared Science Archive, which is operated by the Jet
Propulsion Laboratory, California Institute of Technology, under
contract with the National Aeronautics and Space Administration.}
\end{acknowledgements}

\clearpage

\clearpage

\clearpage

\clearpage

\begin{longtable}{lccccccccc}
\caption{The relevant information of four GMCs.}\label{tbl-total}\\
\endfirsthead
\caption{(continued)} \\
\endhead

\hline \hline
Sources & ID      &    R.A.(J2000)  &   DEC.(J2000)  & Distance  &  Size($\alpha\times\delta$)&  M($H_{2}$)  &   Int. time &   A/B/C      & References    \\

&    (No.)  &   ( $^h$\quad $^m$\quad $^s$)     &   ( $^\circ$\quad$'$\quad$''$)  &  (kpc)  &   (arcmin$^2$) & (M$_{\odot}$)    &       (minutes)     &           &        \\

\hline

W49      &    12   &   19 10 15.25   &    ~09 06 08.4  &  11.4   &      50$\times$50   &  2296910  &   966           &   19/5/2     &   $(1),~(5)$         \\
W3       &    17   &   02 25 38.79   &    ~62 02 22.0  &  1.95   &      70$\times$90   &  71615    &  2370           &  41/19/4     &   $(2),~(5)$         \\
DR21     &    18   &   20 39 01.23   &    ~42 19 33.9  &   3.0   &      60$\times$90   &  444774   &  2742           &  45/34/8     &   $(3),~(5)$         \\
NGC2024  &    27   &   05 41 45.49   &    -01 54 46.8  &  0.415  &      50$\times$100  &  5609     &  3378           &  46/28/2     &   $(4),~(5)$         \\
Total    &         &                 &                 &         &     230$\times$330  &           &  9456           & 151/86/16    &                 \\

\hline
\end{longtable}
\raggedright \emph{References:} ($1$) \citet{gwi92}; ($2$)
\citet{xu06}; ($3$) \citet{cam82}; ($4$)
 \citet{men07}; ($5$)  \citet{bie82}.\\
\emph{Notes:}------ In Column 7 the clump's $H_{2}$-masses
(M($H_{2}$)) are derived from Equation \ref{eq:mh22}. In Column 9,
$A$ is the number of all observational positions toward the sources,
$B$ is the number of H$_{2}$CO absorption lines, and $C$ is the
number of H110$\alpha$ RRLs. Hence, the detection rate is 56.95\%
for H$_{2}$CO absorption lines and 10.60\% for H110$\alpha$ RRLs.

\addtolength{\hoffset}{-0.4in} \setlength{\tabcolsep}{2.8pt}

\clearpage  \scriptsize
\begin{longtable}{cc|cccccccc|cccc}
\caption{The parameters of W49 GMC.}\label{tbl-w49}\\
\endfirsthead
\caption{(continued)} \\
\endhead

\hline \hline
 W49  &            &   H$_{2}$CO          &                &                 &  & & & &                  &   H110$\alpha$     &                 &                  &                 \\
\hline
(1)   & (2) &  (3) &  (4) &  (5)  & (6)  & (7) & (8) & (9) &  (10)  &   (11) &  (12)  &  (13)  & (14)  \\
ID    & Offset($\alpha$, $\delta$) &    Velocity    & Flux   &  $\Delta V$    & Intensity  & $T_{c}$ & $\tau_{app}$ & $N(H_{2}CO)$ &  $N(H_{2})$&   Velocity     &   Flux   &  $\Delta V$  & Intensity      \\
(No.) &   (arcmin)  &  (km s$^{-1}$)      &   (Jy km s$^{-1}$)   &  (km s$^{-1}$)       &  (Jy)  &(K)&&($10^{13}cm^{-2})$&($10^{22}cm^{-2})$         &(km s$^{-1}$)    &  (Jy km s$^{-1}$)    & (km s$^{-1}$)            &  (Jy)          \\
\hline

 10  &   20, 0    & 4.733(0.68)  &    -0.73(0.20)   &   4.61(1.41)  &  -0.14(0.036)  & 0.406  &   0.012  &   0.503   &  0.403   &   N         &               &                &                 \\
     &            & 42.18(0.11)  &    -1.72(0.15)   &   2.37(0.25)  &  -0.68(0.036)  & 0.406  &   0.058  &   1.286   &  1.029   &             &               &                &                 \\
     &            & 45.51(0.32)  &    -0.58(0.13)   &   2.32(0.59)  &  -0.23(0.036)  & 0.406  &   0.019  &   0.418   &  0.334   &             &               &                &                 \\
     &            & 62.09(0.30)  &    -1.75(0.20)   &   5.05(0.66)  &  -0.32(0.036)  & 0.406  &   0.027  &   1.270   &  1.016   &             &               &                &                 \\
 11  &   10, 0    & 12.68(0.69)  &    -2.91(0.52)   &   8.02(1.56)  &  -0.34(0.073)  & 2.060  &   0.013  &   0.978   &  0.782   & 7.52(1.47)  &   3.03(0.44)  &    19.08(2.74) &   0.15(0.039)   \\
     &            & 18.64(0.45)  &    -0.84(0.41)   &   2.94(1.12)  &  -0.26(0.073)  & 2.060  &   0.010  &   0.274   &  0.219   &             &               &                &                 \\
     &            & 41.69(0.09)  &    -3.46(0.22)   &   2.77(0.23)  &  -1.17(0.073)  & 2.060  &   0.045  &   1.181   &  0.945   &             &               &                &                 \\
     &            & 61.50(0.39)  &    -2.84(0.49)   &   5.00(0.77)  &  -0.53(0.073)  & 2.060  &   0.020  &   0.954   &  0.763   &             &               &                &                 \\
     &            & 67.36(0.65)  &    -1.43(0.50)   &   4.50(1.74)  &  -0.29(0.073)  & 2.060  &   0.011  &   0.468   &  0.374   &             &               &                &                 \\
 12  &    0, 0    & 10.49(0.81)  &   -11.51(2.03)   &   9.67(0.90)  &  -1.12(0.099)  & 5.705  &   0.020  &   1.779   &  1.423   & 7.54(0.35)  &  31.39(0.82)  &    27.88(0.85) &   1.06(0.104)   \\
     &            & 15.88(0.16)  &   -13.05(2.03)   &   5.56(0.36)  &  -2.20(0.099)  & 5.705  &   0.039  &   2.028   &  1.622   &             &               &                &                 \\
     &            & 40.86(0.11)  &    -3.67(0.31)   &   2.88(0.30)  &  -1.20(0.099)  & 5.705  &   0.021  &   0.568   &  0.454   &             &               &                &                 \\
     &            & 64.25(0.12)  &    -3.76(0.34)   &   3.09(0.38)  &  -1.15(0.099)  & 5.705  &   0.020  &   0.584   &  0.467   &             &               &                &                 \\
 14  &   20, -10  & 11.99(0.57)  &    -0.26(0.14)   &   1.39(0.71)  &  -0.17(0.040)  & 0.102  &   0.018  &   0.236   &  0.189   &   N         &               &                &                 \\
     &            & 42.36(0.16)  &    -1.37(0.18)   &   2.73(0.47)  &  -0.47(0.040)  & 0.102  &   0.051  &   1.302   &  1.042   &             &               &                &                 \\
 16  &    0, -10  & 12.54(1.92)  &    -1.10(0.41)   &   8.99(4.15)  &  -0.11(0.035)  & 0.414  &   0.009  &   0.766   &  0.613   &   N         &               &                &                 \\
     &            & 15.66(0.30)  &    -0.58(0.23)   &   1.78(0.48)  &  -0.30(0.035)  & 0.414  &   0.025  &   0.417   &  0.333   &             &               &                &                 \\
     &            & 41.93(0.43)  &    -1.14(0.23)   &   4.15(0.92)  &  -0.26(0.035)  & 0.414  &   0.022  &   0.841   &  0.673   &             &               &                &                 \\
     &            & 60.91(0.82)  &    -0.23(0.26)   &   1.70(11.5)  &  -0.13(0.035)  & 0.414  &   0.011  &   0.171   &  0.137   &             &               &                &                 \\
     &            & 63.63(0.70)  &    -0.36(0.23)   &   1.74(0.87)  &  -0.19(0.035)  & 0.414  &   0.016  &   0.257   &  0.206   &             &               &                &                 \\

\hline
\end{longtable}
\raggedright \emph{Notes:}------ Parameters listed about
simultaneously observing H$_{2}$CO absorption line and H110$\alpha$
RRL, and each one of both. The serial number and offset are
indicated in Column one and two corresponding to spectra and color
map. $''N''$ indicates that the corresponding spectra could not be
detected. These intensity data of H110$\alpha$ RRL is not able to
achieve 3$\sigma$ with $''a''$ to line out, so we do not consider
them as signal to analysis.

\clearpage 
\begin{longtable}{cc|cccccccc|cccc}
\caption{The parameters of W3 GMC.}\label{tbl-w3} \\
\endfirsthead
\caption{(continued)} \\
\endhead
\hline \hline
 W3  &            &   H$_{2}$CO          &                &                 &&&&&                   &   H110$\alpha$     &                 &                  &            \\
\hline
(1)   & (2) &  (3) &  (4) &  (5)  & (6)  & (7) & (8) & (9) &  (10)  &   (11) &  (12)  &  (13)  & (14)  \\
ID    & Offset($\alpha$, $\delta$) &    Velocity    & Flux   &  $\Delta V$    & Intensity  & $T_{c}$ & $\tau_{app}$ & $N(H_{2}CO)$ &  $N(H_{2})$&   Velocity     &   Flux   &  $\Delta V$  & Intensity      \\
(No.) &   (arcmin)  &  (km s$^{-1}$)      &   (Jy km s$^{-1}$)   &  (km s$^{-1}$)       &  (Jy)  &(K)&&($10^{13}cm^{-2})$&($10^{22}cm^{-2})$         &(km s$^{-1}$)    &  (Jy km s$^{-1}$)    & (km s$^{-1}$)            &  (Jy)          \\
\hline

 05   &   10, 20    &  -38.73(0.28)   & -1.03(0.13)   &      4.37(0.62)   &     -0.22(0.069) & 0.892 &  0.014 & 0.558 & 0.446   &   -44.15(1.69)  &    3.67(0.50)  &      24.82(4.10)   &    0.13(0.056)$^{a}$    \\
 08   &  -20, 20    &  -35.02(0.58)   & -0.78(0.23)   &      3.90(1.29)   &     -0.18(0.022) & 0.305 &  0.016 & 0.591 & 0.473   &    N            &                &                    &                         \\
 10   &   10, 10    &  -39.37(0.18)   & -2.04(0.27)   &      2.54(0.45)   &     -0.75(0.078) & 2.269 &  0.027 & 0.644 & 0.515   &   -44.66(0.80)  &   11.87(0.69)  &      27.89(1.91)   &    0.40(0.080)          \\
 11   &    0, 10    &  -36.79(0.13)   & -5.17(0.30)   &      4.66(0.33)   &     -1.04(0.104) & 2.784 &  0.032 & 1.419 & 1.135   &   -40.79(0.72)  &   12.54(0.66)  &      27.79(1.74)   &    0.42(0.083)          \\
 12   &  -10, 10    &  -37.32(0.42)   & -0.77(0.16)   &      3.73(0.73)   &     -0.19(0.031) & 0.514 &  0.015 & 0.514 & 0.411   &    N            &                &                    &                         \\
 13   &  -20, 10    &  -37.42(0.26)   & -0.65(0.16)   &      1.70(0.39)   &     -0.36(0.002) & 0.368 &  0.031 & 0.495 & 0.396   &    N            &                &                    &                         \\
 14   &  -30, 10    &  -41.79(0.18)   & -0.43(0.18)   &      1.20(3.01)   &     -0.34(0.001) & 0.304 &  0.031 & 0.346 & 0.277   &    N            &                &                    &                         \\
      &             &  -38.25(0.20)   & -0.73(0.17)   &      1.50(0.68)   &     -0.45(0.001) & 0.304 &  0.041 & 0.576 & 0.461   &                 &                &                    &                         \\
 16   &   10, 0     &  -36.47(0.39)   & -1.03(0.17)   &      4.45(0.95)   &     -0.21(0.035) & 1.659 &  0.009 & 0.385 & 0.308   &   -41.70(0.62)  &    8.68(0.46)  &      24.30(1.59)   &    0.33(0.051)          \\
 17   &    0, 0     &  -36.57(0.09)   & -7.26(0.38)   &      3.79(0.24)   &     -1.80(0.075) & 5.850 &  0.031 & 1.103 & 0.882   &   -41.97(0.42)  &   28.59(0.95)  &      26.12(1.05)   &    1.03(0.121)          \\
 18   &  -10, 0     &  -40.86(0.31)   & -0.64(0.18)   &      2.33(0.81)   &     -0.26(0.009) & 0.957 &  0.016 & 0.340 & 0.272   &   -38.81(2.50)  &    3.09(0.56)  &      26.58(5.07)   &    0.11(0.065)$^{a}$    \\
      &             &  -37.15(0.16)   & -1.08(0.16)   &      1.86(0.38)   &     -0.54(0.009) & 0.957 &  0.033 & 0.569 & 0.455   &                 &                &                    &                         \\
 19   &  -20, 0     &  -39.54(0.12)   & -1.96(0.16)   &      2.91(0.28)   &     -0.63(0.048) & 0.475 &  0.051 & 1.390 & 1.112   &    N            &                &                    &                         \\
 20   &  -30, 0     &  -39.78(0.20)   & -1.19(0.22)   &      2.24(0.54)   &     -0.49(0.070) & 0.375 &  0.042 & 0.889 & 0.711   &    N            &                &                    &                         \\
 22   &   10, -10   &  -43.49(0.30)   & -1.55(0.23)   &      4.25(0.82)   &     -0.34(0.066) & 1.013 &  0.020 & 0.790 & 0.632   &   -40.71(2.10)  &    3.59(0.54)  &      26.59(3.98)   &    0.12(0.063)$^{a}$    \\
 23   &    0, -10   &   N             &               &                   &                  & 0.844 &        &       &         &   -38.45(1.84)  &    2.66(0.71)  &      14.25(5.48)   &    0.17(0.080)$^{a}$    \\
 24   &  -10, -10   &  -40.93(0.29)   & -0.61(0.26)   &      1.20(2.09)   &     -0.48(0.016) & 0.660 &  0.034 & 0.385 & 0.308   &   -39.34(1.91)  &    2.29(0.42)  &      20.81(4.50)   &    0.10(0.062)$^{a}$    \\
      &             &  -38.89(0.49)   & -1.09(0.32)   &      3.15(0.82)   &     -0.32(0.016) & 0.660 &  0.023 & 0.670 & 0.536   &                 &                &                    &                         \\
 25   &  -20, -10   &  -39.61(0.27)   & -1.55(0.31)   &      3.00(0.87)   &     -0.48(0.049) & 0.582 &  0.036 & 1.011 & 0.809   &   -41.93(2.42)  &    2.58(0.53)  &      21.30(4.02)   &    0.11(0.078)$^{a}$    \\
 33   &   20, -30   &  -46.30(0.18)   & -2.38(0.26)   &      3.14(0.38)   &     -0.71(0.055) & 1.000 &  0.042 & 1.241 & 0.993   &   -46.05(4.29)  &    3.19(0.79)  &      32.19(8.33)   &    0.09(0.077)$^{a}$    \\
 34   &   10, -30   &  -45.04(0.14)   & -2.66(0.25)   &      3.04(0.34)   &     -0.82(0.054) & 0.658 &  0.059 & 1.688 & 1.351   &    N            &                &                    &                         \\
 37   &   20, -40   &  -45.14(0.14)   & -2.43(0.20)   &      3.32(0.31)   &     -0.68(0.061) & 0.948 &  0.041 & 1.290 & 1.032   &    N            &                &                    &                         \\
 38   &   10, -40   &  -45.20(0.16)   & -0.72(0.15)   &      1.64(0.64)   &     -0.41(0.016) & 0.646 &  0.029 & 0.452 & 0.362   &    N            &                &                    &                         \\

\hline
\end{longtable}
\raggedright \emph{Notes:}------ Parameters listed
about simultaneously observing H$_{2}$CO absorption line and
H110$\alpha$ RRL, and each one of both. The serial number and offset
are indicated in Column one and two corresponding to spectra and
color map. $''N''$ indicates that the corresponding spectra could
not be detected. These intensity data of H110$\alpha$ RRL is not
able to achieve 3$\sigma$ with $''a''$ to line out, so we do not
consider them as signal to analysis.

\clearpage 
\begin{longtable}{cc|cccccccc|cccc}
\caption{The parameters of DR21/W75 GMC.}\label{tbl-dr21}\\
\endfirsthead
\caption{(continued)} \\
\hline
\endhead

\hline \hline
 DR21  &           &   H$_{2}$CO          &                &                 &  &&&&                 &   H110$\alpha$     &                 &                  &     \\
\hline
(1)   & (2) &  (3) &  (4) &  (5)  & (6)  & (7) & (8) & (9) &  (10)  &   (11) &  (12)  &  (13)  & (14)  \\
ID    & Offset($\alpha$, $\delta$) &    Velocity    & Flux   &  $\Delta V$    & Intensity  & $T_{c}$ & $\tau_{app}$ & $N(H_{2}CO)$ &  $N(H_{2})$&   Velocity     &   Flux   &  $\Delta V$  & Intensity      \\
(No.) &   (arcmin)  &  (km s$^{-1}$)      &   (Jy km s$^{-1}$)   &  (km s$^{-1}$)       &  (Jy)  &(K)&&($10^{13}cm^{-2})$&($10^{22}cm^{-2})$         &(km s$^{-1}$)    &  (Jy km s$^{-1}$)    & (km s$^{-1}$)            &  (Jy)          \\
\hline

 04  &  -10, 30   &   0.74(0.31)   &  -0.61(0.16)  &     2.10(0.66) &   -0.27(0.019)  &  0.680  & 0.019  & 0.372 & 0.297         &    N           &             &                     &                     \\
 05  &  -20, 30   &   3.43(0.46)   &  -1.26(0.24)  &     4.84(1.10) &   -0.24(0.025)  &  0.402  & 0.020  & 0.913 & 0.730         &    N           &             &                     &                     \\
 07  &   10, 20   &  13.97(0.43)   &  -0.69(0.19)  &     2.09(0.84) &   -0.31(0.052)  &  0.393  & 0.026  & 0.514 & 0.411         &    N           &             &                     &                     \\
 08  &    0, 20   &  -0.60(0.24)   &  -0.73(0.16)  &     2.08(0.54) &   -0.33(0.019)  &  0.476  & 0.026  & 0.514 & 0.411         &   2.93(2.00)   &  2.09(0.47) &       17.76(4.62)   &   0.11(0.055)$^{a}$ \\
     &            &  11.90(0.26)   &  -0.63(0.18)  &     2.13(0.85) &   -0.28(0.019)  &  0.476  & 0.022  & 0.445 & 0.356         &                &             &                     &                     \\
     &            &  14.86(0.15)   &  -0.39(0.13)  &     1.20(13.1) &   -0.30(0.019)  &  0.476  & 0.024  & 0.269 & 0.215         &                &             &                     &                     \\
     &            &  17.68(0.27)   &  -0.34(0.11)  &     1.20(5.50) &   -0.26(0.019)  &  0.476  & 0.021  & 0.233 & 0.186         &                &             &                     &                     \\
 09  &  -10, 20   &   1.01(0.16)   &  -2.94(0.30)  &     3.16(0.35) &   -0.87(0.034)  &  0.434  & 0.073  & 2.168 & 1.735         &    N           &             &                     &                     \\
 11  &   20, 10   &  10.82(0.33)   &  -0.85(0.20)  &     2.65(0.63) &   -0.30(0.027)  &  0.406  & 0.025  & 0.624 & 0.499         &   4.70(1.79)   &  1.40(0.49) &       10.35(4.29)   &   0.12(0.086)$^{a}$ \\
 12  &   10, 10   &  11.23(0.29)   &  -1.11(0.21)  &     3.12(0.73) &   -0.33(0.017)  &  0.467  & 0.026  & 0.775 & 0.620         &    N           &             &                     &                     \\
     &            &  15.16(0.68)   &  -0.27(0.16)  &     2.06(1.19) &   -0.12(0.017)  &  0.467  & 0.010  & 0.185 & 0.148         &                &             &                     &                     \\
 13  &    0, 10   &   0.00(0.11)   &  -2.72(0.18)  &     3.45(0.26) &   -0.74(0.026)  &  0.628  & 0.054  & 1.757 & 1.406         &    N           &             &                     &                     \\
     &            &  12.33(0.18)   &  -0.73(0.11)  &     1.44(0.50) &   -0.47(0.026)  &  0.628  & 0.034  & 0.461 & 0.369         &                &             &                     &                     \\
 14  &  -10, 10   &    N           &               &                &                 &  0.591  &        &       &               &   5.49(1.64)   &  2.52(0.49) &       15.19(2.89)   &   0.15(0.097)$^{a}$ \\
 16  &   20, 0    &    N           &               &                &                 &  0.931  &        &       &               &   6.42(3.59)   &  3.07(0.79) &       26.44(10.2)   &   0.10(0.070)$^{a}$ \\
 17  &   10, 0    &  -0.24(0.24)   &  -1.08(0.19)  &     2.86(0.63) &   -0.35(0.058)  &  1.136  & 0.019  & 0.516 & 0.413         &   4.73(0.96)   &  4.10(0.46) &       17.85(2.46)   &   0.21(0.069)       \\
     &            &   6.24(0.33)   &  -0.49(0.18)  &     2.09(1.12) &   -0.22(0.058)  &  1.136  & 0.012  & 0.236 & 0.189         &                &             &                     &                     \\
     &            &  10.93(0.35)   &  -0.92(0.18)  &     2.91(0.64) &   -0.29(0.058)  &  1.136  & 0.016  & 0.434 & 0.347         &                &             &                     &                     \\
     &            &  21.82(0.51)   &  -0.62(0.19)  &     3.21(1.07) &   -0.18(0.058)  &  1.136  & 0.010  & 0.296 & 0.237         &                &             &                     &                     \\
 18  &    0, 0    &   0.26(0.03)   & -13.76(0.27)  &     3.70(0.07) &   -3.49(0.164)  &  3.128  & 0.103  & 3.590 & 2.872         &   2.14(1.32)   &  9.74(0.73) &       34.96(3.10)   &   0.26(0.085)       \\
     &            &  10.91(0.10)   &  -2.69(0.22)  &     2.33(0.20) &   -1.08(0.164)  &  3.128  & 0.031  & 0.675 & 0.540         &                &             &                     &                     \\
 19  &  -10, 0    &   1.63(0.26)   &  -0.88(0.16)  &     2.80(0.57) &   -0.29(0.011)  &  0.996  & 0.017  & 0.447 & 0.358         &   8.11(1.64)   &  3.20(0.53) &       17.51(3.27)   &   0.17(0.069)$^{a}$ \\
     &            &  12.18(0.28)   &  -0.54(0.15)  &     2.14(0.79) &   -0.23(0.011)  &  0.996  & 0.013  & 0.271 & 0.217         &                &             &                     &                     \\
 20  &  -20, 0    &   1.70(0.07)   &  -0.77(0.12)  &     1.20(2.55) &   -0.60(0.000)  &  0.772  & 0.040  & 0.452 & 0.362         &   9.73(2.22)   &  3.54(0.61) &       25.98(5.34)   &   0.12(0.054)$^{a}$ \\
 21  &  -30, 0    &   1.77(0.34)   &  -0.61(0.25)  &     2.43(1.56) &   -0.23(0.010)  &  0.724  & 0.016  & 0.356 & 0.285         &    N           &             &                     &                     \\
     &            &   8.22(0.22)   &  -0.81(0.17)  &     1.78(0.40) &   -0.43(0.010)  &  0.724  & 0.029  & 0.491 & 0.393         &                &             &                     &                     \\
 22  &   20, -10  &  -2.70(0.74)   &  -0.48(0.22)  &     2.99(1.25) &   -0.15(0.021)  &  1.455  & 0.007  & 0.200 & 0.160         &   6.96(1.18)   &  5.88(0.57) &       23.66(2.42)   &   0.23(0.067)       \\
     &            &   6.49(0.16)   &  -1.27(0.21)  &     2.26(0.53) &   -0.52(0.021)  &  1.455  & 0.025  & 0.528 & 0.423         &                &             &                     &                     \\
     &            &  10.60(0.18)   &  -1.13(0.19)  &     2.06(0.35) &   -0.51(0.021)  &  1.455  & 0.024  & 0.472 & 0.378         &                &             &                     &                     \\
 23  &   10, -10  &   6.44(0.10)   &  -1.78(0.15)  &     2.17(0.17) &   -0.77(0.049)  &  1.455  & 0.037  & 0.756 & 0.605         &   8.86(0.71)   &  7.48(0.50) &       21.37(1.62)   &   0.32(0.064)       \\
     &            &  11.55(0.13)   &  -1.18(0.18)  &     1.90(0.42) &   -0.58(0.049)  &  1.455  & 0.028  & 0.496 & 0.397         &                &             &                     &                     \\
 24  &    0, -10  &   0.42(0.25)   &  -1.40(0.19)  &     3.64(0.50) &   -0.36(0.045)  &  1.156  & 0.020  & 0.669 & 0.535         &   8.14(1.07)   &  4.33(0.46) &       20.52(2.67)   &   0.19(0.063)       \\
     &            &   6.36(0.16)   &  -0.86(0.14)  &     1.81(0.38) &   -0.44(0.045)  &  1.156  & 0.024  & 0.408 & 0.326         &                &             &                     &                     \\
     &            &  11.21(0.11)   &  -1.11(0.12)  &     1.44(0.47) &   -0.72(0.045)  &  1.156  & 0.040  & 0.535 & 0.428         &                &             &                     &                     \\
 25  &  -10, -10  &   6.04(0.09)   &  -1.93(0.17)  &     2.06(0.19) &   -0.88(0.030)  &  1.148  & 0.049  & 0.943 & 0.754         &   6.04(1.20)   &  3.68(0.46) &       20.14(3.06)   &   0.17(0.063)$^{a}$ \\
 26  &  -20, -10  &   6.99(0.32)   &  -1.00(0.22)  &     2.76(0.66) &   -0.34(0.053)  &  0.922  & 0.021  & 0.538 & 0.430         &   4.51(2.09)   &  2.80(0.57) &       19.54(3.99)   &   0.13(0.072)$^{a}$ \\
     &            &  18.41(0.26)   &  -1.25(0.22)  &     2.74(0.52) &   -0.43(0.053)  &  0.922  & 0.026  & 0.677 & 0.542         &                &             &                     &                     \\
 27  &  -30, -10  &   7.88(0.39)   &  -0.56(0.22)  &     1.21(25.2) &   -0.43(0.040)  &  0.649  & 0.031  & 0.349 & 0.279         &    N           &             &                     &                     \\
     &            &  17.17(0.22)   &  -2.15(0.25)  &     3.69(0.47) &   -0.54(0.040)  &  0.649  & 0.039  & 1.343 & 1.074         &                &             &                     &                     \\
 28  &   20, -20  &  -4.09(0.44)   &  -1.16(0.32)  &     2.89(0.94) &   -0.37(0.027)  &  1.964  & 0.015  & 0.396 & 0.317         &   0.14(0.68)   &  7.20(0.71) &       14.87(1.97)   &   0.45(0.095)       \\
     &            &   6.11(0.36)   &  -0.60(0.21)  &     1.20(1.11) &   -0.46(0.027)  &  1.964  & 0.018  & 0.205 & 0.164         &                &             &                     &                     \\
     &            &  11.14(0.20)   &  -1.23(0.28)  &     1.98(0.59) &   -0.58(0.027)  &  1.964  & 0.023  & 0.427 & 0.342         &                &             &                     &                     \\
 29  &   10, -20  &   5.87(0.16)   &  -1.28(0.21)  &     1.75(0.25) &   -0.68(0.013)  &  1.577  & 0.031  & 0.511 & 0.409         &   3.59(1.56)   &  6.41(0.90) &       23.39(4.18)   &   0.25(0.080)       \\
     &            &  11.18(0.14)   &  -2.07(0.26)  &     2.31(0.36) &   -0.83(0.013)  &  1.577  & 0.038  & 0.827 & 0.661         &                &             &                     &                     \\
 30  &    0, -20  &   6.17(0.11)   &  -2.73(0.33)  &     2.03(0.32) &   -1.26(0.111)  &  1.111  & 0.072  & 1.369 & 1.095         &    N           &             &                     &                     \\
     &            &  11.15(0.25)   &  -1.14(0.32)  &     1.77(0.89) &   -0.60(0.111)  &  1.111  & 0.034  & 0.558 & 0.446         &                &             &                     &                     \\
 31  &  -10, -20  &   6.45(0.04)   &  -3.91(0.23)  &     1.47(0.42) &   -2.49(0.105)  &  0.682  & 0.188  & 2.603 & 2.083         &    N           &             &                     &                     \\
 32  &  -20, -20  &   6.39(0.09)   &  -3.36(0.30)  &     2.28(0.28) &   -1.38(0.126)  &  0.674  & 0.101  & 2.155 & 1.724         &    N           &             &                     &                     \\
 33  &  -30, -20  &   5.36(0.15)   &  -0.44(0.16)  &     1.20(3.14) &   -0.34(0.051)  &  0.523  & 0.026  & 0.296 & 0.237         &    N           &             &                     &                     \\
     &            &   9.57(0.30)   &  -0.97(0.21)  &     2.58(0.59) &   -0.35(0.051)  &  0.523  & 0.027  & 0.655 & 0.524         &                &             &                     &                     \\
     &            &  16.97(0.58)   &  -0.78(0.23)  &     3.56(0.90) &   -0.20(0.051)  &  0.523  & 0.015  & 0.514 & 0.411         &                &             &                     &                     \\
 34  &   20, -30  &  -2.89(0.20)   &  -1.62(0.25)  &     2.51(0.46) &   -0.60(0.115)  &  0.404  & 0.051  & 1.199 & 0.960         &  -2.69(1.15)   &  7.54(0.88) &       21.89(3.56)   &   0.32(0.085)       \\
     &            &  12.16(0.39)   &  -1.09(0.24)  &     2.71(0.60) &   -0.37(0.115)  &  0.404  & 0.031  & 0.791 & 0.633         &                &             &                     &                     \\
 35  &   10, -30  &  -2.49(0.29)   &  -0.62(0.15)  &     2.08(0.43) &   -0.28(0.010)  &  1.064  & 0.016  & 0.310 & 0.248         &    N           &             &                     &                     \\
     &            &   5.64(0.24)   &  -0.59(0.14)  &     1.73(0.55) &   -0.32(0.010)  &  1.064  & 0.018  & 0.295 & 0.236         &                &             &                     &                     \\
     &            &  12.31(0.29)   &  -0.24(0.12)  &     1.20(0.68) &   -0.19(0.010)  &  1.064  & 0.011  & 0.121 & 0.097         &                &             &                     &                     \\
 36  &    0, -30  &   6.36(0.44)   &  -0.62(0.17)  &     2.33(0.92) &   -0.25(0.067)  &  0.897  & 0.015  & 0.337 & 0.270         &    N           &             &                     &                     \\
 37  &  -10, -30  &   6.08(0.23)   &  -0.95(0.23)  &     1.71(0.46) &   -0.52(0.009)  &  1.507  & 0.024  & 0.391 & 0.313         &    N           &             &                     &                     \\
 38  &  -20, -30  &   8.12(0.57)   &  -1.76(0.36)  &     6.09(1.42) &   -0.27(0.025)  &  0.890  & 0.017  & 0.957 & 0.765         &    N           &             &                     &                     \\
     &            &  10.66(0.49)   &  -0.27(0.13)  &     1.20(0.73) &   -0.21(0.025)  &  0.890  & 0.013  & 0.146 & 0.117         &                &             &                     &                     \\
     &            &  15.10(0.11)   &  -0.56(0.13)  &     1.20(1.55) &   -0.43(0.025)  &  0.890  & 0.027  & 0.302 & 0.241         &                &             &                     &                     \\
 39  &  -30, -30  &   8.71(0.25)   &  -1.44(0.28)  &     2.61(0.63) &   -0.51(0.037)  &  0.606  & 0.038  & 0.921 & 0.737         &    N           &             &                     &                     \\
 43  &  -10, -40  &   8.18(0.24)   &  -1.15(0.26)  &     2.08(0.53) &   -0.52(0.015)  &  0.129  & 0.055  & 1.074 & 0.859         &    N           &             &                     &                     \\
 44  &  -20, -40  &   0.11(0.24)   &  -0.73(0.21)  &     1.68(0.61) &   -0.41(0.065)  &  0.000  & 0.049  & 0.769 & 0.616         &  -2.61(2.80)   &  3.03(0.75) &       22.15(7.36)   &   0.12(0.075)$^{a}$ \\
     &            &   8.59(0.24)   &  -1.41(0.24)  &     2.63(0.54) &   -0.50(0.065)  &  0.000  & 0.060  & 1.477 & 1.182         &                &             &                     &                     \\
 45  &  -30, -40  &   0.18(0.48)   &  -0.50(0.16)  &     2.80(0.74) &   -0.16(0.036)  &  0.094  & 0.017  & 0.450 & 0.360         &    N           &             &                     &                     \\
     &            &   8.91(0.09)   &  -0.57(0.12)  &     1.20(5.32) &   -0.45(0.036)  &  0.094  & 0.049  & 0.551 & 0.441         &                &             &                     &                     \\

\hline
\end{longtable}
\raggedright \emph{Notes:}------ Parameters listed about
simultaneously observing H$_{2}$CO absorption line and H110$\alpha$
RRL, and each one of both. The serial number and offset are
indicated in Column one and two corresponding to spectra and color
map. $''N''$ indicates that the corresponding spectra could not be
detected. These intensity data of H110$\alpha$ RRL is not able to
achieve 3$\sigma$ with $''a''$ to line out, so we do not consider
them as signal to analysis.

\clearpage

\begin{longtable}{cc|cccccccc|cccc}

\caption{The parameters of NGC2024/2023 GMC.}\label{tbl-ngc2024}\\
\endfirsthead
\caption{(continued)} \\
\hline
\endhead

\hline\hline
 NGC2024  &            &   H$_{2}$CO          &                &                 &   &&&&                &   H110$\alpha$     &                 &                  &    \\
\hline
(1)   & (2) &  (3) &  (4) &  (5)  & (6)  & (7) & (8) & (9) &  (10)  &   (11) &  (12)  &  (13)  & (14)  \\
ID    & Offset($\alpha$, $\delta$) &    Velocity    & Flux   &  $\Delta V$    & Intensity  & $T_{c}$ & $\tau_{app}$ & $N(H_{2}CO)$ &  $N(H_{2})$&   Velocity     &   Flux   &  $\Delta V$  & Intensity      \\
(No.) &   (arcmin)  &  (km s$^{-1}$)      &   (Jy km s$^{-1}$)   &  (km s$^{-1}$)       &  (Jy)  &(K)&&($10^{13}cm^{-2})$&($10^{22}cm^{-2})$         &(km s$^{-1}$)    &  (Jy km s$^{-1}$)    & (km s$^{-1}$)            &  (Jy)          \\
\hline

  01  &   20, 50   &     5.40(0.35)  &   -0.88(0.17)   &    3.51(0.64)  &    -0.23(0.032) & 0.058 & 0.026 & 0.842 & 0.674  &   N          &              &                   &                     \\
      &            &    10.99(0.26)  &   -0.48(0.13)   &    1.80(0.61)  &    -0.24(0.032) & 0.058 & 0.027 & 0.451 & 0.361  &              &              &                   &                     \\
  03  &    0, 50   &     7.49(0.29)  &   -0.19(0.07)   &    1.20(1.74)  &    -0.14(0.001) & 0.024 & 0.016 & 0.180 & 0.144  &   N          &              &                   &                     \\
      &            &    12.27(0.21)  &   -0.29(0.08)   &    1.39(0.42)  &    -0.19(0.001) & 0.024 & 0.022 & 0.284 & 0.227  &              &              &                   &                     \\
  05  &   20, 40   &     4.01(0.14)  &   -2.08(0.25)   &    2.33(0.39)  &    -0.83(0.092) & 0.095 & 0.092 & 2.016 & 1.612  &   N          &              &                   &                     \\
      &            &    10.68(0.14)  &   -1.57(0.20)   &    2.04(0.26)  &    -0.72(0.092) & 0.095 & 0.079 & 1.521 & 1.217  &              &              &                   &                     \\
  06  &   10, 40   &     4.13(0.19)  &   -0.75(0.18)   &    1.91(0.72)  &    -0.37(0.029) & 0.060 & 0.041 & 0.742 & 0.594  &   N          &              &                   &                     \\
      &            &    10.61(0.14)  &   -1.46(0.18)   &    2.27(0.30)  &    -0.60(0.029) & 0.060 & 0.068 & 1.449 & 1.159  &              &              &                   &                     \\
  07  &    0, 40   &    11.28(0.26)  &   -0.80(0.15)   &    2.57(0.51)  &    -0.29(0.050) & 0.038 & 0.033 & 0.796 & 0.637  &   N          &              &                   &                     \\
  08  &  -10, 40   &    11.43(0.08)  &   -1.31(0.15)   &    1.82(0.36)  &    -0.67(0.002) & 0.034 & 0.078 & 1.337 & 1.070  &   N          &              &                   &                     \\
  10  &   20, 30   &     5.17(0.49)  &   -1.15(0.22)   &    4.54(1.10)  &    -0.23(0.028) & 0.125 & 0.024 & 1.025 & 0.820  &   N          &              &                   &                     \\
      &            &    11.47(0.27)  &   -0.73(0.25)   &    1.20(34.1)  &    -0.56(0.028) & 0.125 & 0.059 & 0.671 & 0.537  &              &              &                   &                     \\
  11  &   10, 30   &     5.28(0.52)  &   -0.55(0.17)   &    2.66(1.18)  &    -0.19(0.040) & 0.125 & 0.020 & 0.495 & 0.396  &   N          &              &                   &                     \\
      &            &    11.23(0.27)  &   -1.11(0.18)   &    3.16(0.58)  &    -0.33(0.040) & 0.125 & 0.035 & 1.028 & 0.823  &              &              &                   &                     \\
  12  &    0, 30   &    11.79(0.31)  &   -1.14(0.20)   &    2.91(0.69)  &    -0.36(0.099) & 0.049 & 0.041 & 1.111 & 0.889  &   N          &              &                   &                     \\
  13  &  -10, 30   &    11.46(0.08)  &   -0.69(0.12)   &    1.20(1.29)  &    -0.53(0.000) & 0.092 & 0.058 & 0.654 & 0.523  &   N          &              &                   &                     \\
  16  &   10, 20   &    11.48(0.13)  &   -0.49(0.12)   &    1.77(1.21)  &    -0.26(0.015) & 0.180 & 0.026 & 0.431 & 0.345  &   N          &              &                   &                     \\
      &            &    14.74(0.14)  &   -0.44(0.09)   &    1.20(0.55)  &    -0.34(0.015) & 0.180 & 0.034 & 0.383 & 0.307  &              &              &                   &                     \\
  17  &    0, 20   &    11.79(0.16)  &   -1.27(0.17)   &    2.27(0.34)  &    -0.52(0.024) & 0.201 & 0.052 & 1.100 & 0.880  &   N          &              &                   &                     \\
  18  &  -10, 20   &    11.11(0.40)  &   -0.53(0.17)   &    2.41(0.97)  &    -0.20(0.023) & 0.001 & 0.023 & 0.531 & 0.425  &   N          &              &                   &                     \\
  21  &   10, 10   &     7.51(0.15)  &   -0.78(0.12)   &    1.73(0.44)  &    -0.42(0.005) & 0.300 & 0.038 & 0.621 & 0.497  &   N          &              &                   &                     \\
  22  &    0, 10   &     6.67(0.23)  &   -0.62(0.17)   &    2.02(0.74)  &    -0.29(0.008) & 0.783 & 0.019 & 0.362 & 0.289  &  1.08(1.28)  &   3.59(0.45) &      20.38(2.87)  &    0.16(0.067)$^{a}$\\
      &            &    11.26(0.08)  &   -2.02(0.14)   &    2.27(0.22)  &    -0.83(0.008) & 0.783 & 0.056 & 1.185 & 0.948  &              &              &                   &                     \\
  23  &  -10, 10   &    11.37(0.29)  &   -0.79(0.15)   &    2.80(0.57)  &    -0.26(0.018) & 0.211 & 0.025 & 0.664 & 0.531  &   N          &              &                   &                     \\
  26  &   10, 0    &    12.19(0.66)  &   -0.85(0.29)   &    2.72(1.66)  &    -0.29(0.081) & 0.485 & 0.023 & 0.586 & 0.469  &   N          &              &                   &                     \\
  27  &    0, 0    &    11.21(0.02)  &  -10.49(0.27)   &    2.18(0.08)  &    -4.51(0.292) & 5.619 & 0.082 & 1.687 & 1.350  &  4.69(0.18)  &  28.03(0.47) &      22.27(0.44)  &    1.18(0.080)      \\
  28  &  -10, 0    &    12.53(0.40)  &   -1.11(0.19)   &    4.35(0.75)  &    -0.24(0.044) & 0.603 & 0.018 & 0.716 & 0.573  &  5.65(1.24)  &   3.69(0.39) &      22.77(2.88)  &    0.15(0.049)      \\
  30  &   20, -10  &    12.68(0.12)  &   -1.02(0.16)   &    1.91(0.51)  &    -0.50(0.020) & 0.254 & 0.047 & 0.850 & 0.680  &   N          &              &                   &                     \\
  31  &   10, -10  &    12.82(0.17)  &   -1.14(0.16)   &    2.42(0.41)  &    -0.44(0.049) & 0.305 & 0.040 & 0.908 & 0.726  &   N          &              &                   &                     \\
  32  &    0, -10  &    13.89(0.16)  &   -1.17(0.16)   &    2.15(0.38)  &    -0.51(0.041) & 0.392 & 0.043 & 0.878 & 0.702  &   N          &              &                   &                     \\
  33  &  -10, -10  &    13.48(0.24)  &   -0.71(0.15)   &    1.51(0.68)  &    -0.44(0.004) & 0.324 & 0.039 & 0.558 & 0.446  &   N          &              &                   &                     \\
  36  &   10, -20  &    13.01(0.28)  &   -0.55(0.12)   &    2.28(0.47)  &    -0.22(0.023) & 0.249 & 0.021 & 0.442 & 0.354  &   N          &              &                   &                     \\
  37  &    0, -20  &    12.66(0.10)  &   -1.83(0.15)   &    2.26(0.24)  &    -0.76(0.102) & 0.307 & 0.070 & 1.483 & 1.186  &   N          &              &                   &                     \\
  38  &  -10, -20  &    12.40(0.19)  &   -0.47(0.14)   &    1.20(28.9)  &    -0.36(0.000) & 0.314 & 0.032 & 0.364 & 0.291  &   N          &              &                   &                     \\
  41  &    0, -30  &    12.34(0.22)  &   -1.56(0.23)   &    2.98(0.51)  &    -0.49(0.071) & 0.313 & 0.044 & 1.240 & 0.992  &   N          &              &                   &                     \\
  42  &  -10, -30  &    13.13(0.29)  &   -0.55(0.15)   &    1.80(0.44)  &    -0.28(0.004) & 0.371 & 0.024 & 0.406 & 0.325  &   N          &              &                   &                     \\

\hline
\end{longtable}
\raggedright \emph{Notes:}------ Parameters listed
about simultaneously observing H$_{2}$CO absorption line and
H110$\alpha$ RRL, and each one of both. The serial number and offset
are indicated in Column one and two corresponding to spectra and
color map. $''N''$ indicates that the corresponding spectra could
not be detected. These intensity data of H110$\alpha$ RRL is not
able to achieve 3$\sigma$ with $''a''$ to line out, so we do not
consider them as signal to analysis.

\end{document}